\documentclass{aastex61}
\received{2017 May 26}
\revised{2017 December 4}
\accepted{2017 December 20}

\shorttitle{Ages of the Bulge Globular Clusters NGC\,6522 and NGC\,6626}
\shortauthors{Kerber et al.}

\begin{document}

\title{Ages of the Bulge globular clusters NGC\,6522 and NGC\,6626 (M28)
  from HST Proper-motion-cleaned Color-Magnitude Diagrams\footnote{Based on observations with the NASA/ESA {\it Hubble Space Telescope}, obtained at  the Space Telescope Science Institute.}
}

\correspondingauthor{Leandro Kerber}
\email{lokerber@uesc.br}

\author[0000-0002-7435-8748]{L. O. Kerber}
\affil{Universidade de S\~ao Paulo, IAG, Rua do Mat\~ao 1226,
Cidade Universit\'aria, S\~ao Paulo 05508-900, Brazil}
\affil{
Universidade Estadual de Santa Cruz, DCET, Rodovia Jorge Amado km 16, 
Ilh\'eus 45662-000, Bahia, Brazil}

\author[0000-0003-1149-3659]{D. Nardiello}
\affil{Universit\`a di Padova, Dipartimento di Astronomia, Vicolo
 dell'Osservatorio 5, I-35122 Padova, Italy}
\affil{INAF-Osservatorio Astronomico di Padova, Vicolo dell'Osservatorio 5,
I-35122 Padova, Italy}

\author{S. Ortolani}
\affil{Universit\`a di Padova, Dipartimento di Astronomia, Vicolo
 dell'Osservatorio 5, I-35122 Padova, Italy}
\affil{INAF-Osservatorio Astronomico di Padova, Vicolo dell'Osservatorio 5,
I-35122 Padova, Italy}

\author{B. Barbuy}
\affil{Universidade de S\~ao Paulo, IAG, Rua do Mat\~ao 1226,
Cidade Universit\'aria, S\~ao Paulo 05508-900, Brazil}

\author{E. Bica}
\affil{Universidade Federal do Rio Grande do Sul, 
Departamento de Astronomia, CP 15051, Porto Alegre 91501-970, Brazil}

\author[0000-0001-5870-3735]{S. Cassisi}
\affil{Osservatorio Astronomico d'Abruzzo, 
Via M. Maggini sn., I-64100 Teramo, Italy} 


\author{M. Libralato}
\affil{Universit\`a di Padova, Dipartimento di Astronomia, Vicolo
 dell'Osservatorio 5, I-35122 Padova, Italy}
\affil{INAF-Osservatorio Astronomico di Padova, Vicolo dell'Osservatorio 5,
  I-35122 Padova, Italy}
\affil{Space Telescope Science Institute, 3700 San Martin Drive, Baltimore, MD 21218, USA}

\author{R. G. Vieira}
\affil{Universidade de S\~ao Paulo, IAG, Rua do Mat\~ao 1226,
Cidade Universit\'aria, S\~ao Paulo 05508-900, Brazil}



\begin{abstract}
  Bulge globular clusters (GCs) with metallicities [Fe/H] $\lesssim -1.0$ and blue horizontal branches
  are candidates to harbor the oldest populations in the Galaxy.
  Based on the analysis of \textit{HST} proper-motion-cleaned color-magnitude diagrams in filters
  F435W and F625W, we determine physical parameters for the old bulge globular clusters NGC\,6522 and
  NGC\,6626 (M28), both with well-defined blue horizontal branches.
  We compare these results with similar data for the inner halo cluster NGC\,6362.
  These clusters have similar metallicities ($-1.3 \leq$ [Fe/H] $\leq -1.0$)
  obtained from high resolution  spectroscopy.
  We derive ages, distance moduli, and reddening values by means of statistical comparisons between observed 
  and synthetic fiducial lines employing likelihood statistics and the Markov chain Monte Carlo method.
  The synthetic fiducial lines were generated using $\alpha$-enhanced BaSTI and Dartmouth stellar
  evolutionary models, adopting both canonical ($Y \sim$ 0.25) and enhanced ($Y\sim$ 0.30-0.33)
  helium abundances.
  RR Lyrae stars were employed to determine the HB magnitude level,
  providing an independent indicator to constrain the apparent distance modulus and the helium enhancement.
  The shape of the observed fiducial line could be compatible with some helium enhancement for
  NGC\,6522 and NGC\,6626, but the average magnitudes of RR Lyrae stars tend to rule out this hypothesis.
  Assuming canonical helium abundances, BaSTI and Dartmouth models indicate that all three clusters are coeval,
  with ages between $\sim$ 12.5 and 13.0 Gyr.
  The present study also reveals that NGC\,6522 has at least two stellar populations,
  since its CMD shows a significantly wide subgiant branch compatible with 14\%$\pm$2\% and 86\%$\pm$5\%
  for first and second generations, respectively.
\end{abstract}




\section{Introduction} 
\label{intro}

The bulk of the Milky Way (MW) bulge stars show a lower-metallicity end at 
$-1.5 \lesssim$ [Fe/H] $\lesssim -1.0$ (e.g., \citealp{Rojas-Arriagada+14,Rojas-Arriagada+17};
\citealp{Zoccali+17}, and references therein).
These relatively high lower-end metallicities
 are expected from an early fast  chemical enrichment in the 
central parts of the Galaxy,
 as modeled  by, e.g., \citet{Cescutti+08}.
From a selection of bulge globular clusters (GCs) presented by
\citet{Bica+16}, it was shown that their
 metallicity distribution also has a peak at [Fe/H]$\sim -1.0$. 
A subclass of these clusters show a
 blue horizontal branch (BHB), making these combined
characteristics to be indicative, in principle, of a very old age 
\citep{Lee+94}.
Furthermore, \citet{Marin-Franch+09} demonstrated that the GCs with
galactocentric distance ($R_{\rm{GC}}$) less than 10 kpc follow a
flat age-metallicity relation with an age dispersion of only $\sim$ 5\%,
reinforcing the idea that a very low metallicity is not a restrictive
prior condition to find stellar relics in the bulge.

Our main targets in this work, NGC\,6522 and  NGC\,6626 (M28),
are both bulge GCs with similar metallicities ($-1.3 \leq$ [Fe/H] $\leq -1.0$)
and BHBs and are both affected by field contamination and high absorption ($A_{\rm{V}}$ $>$ 1.20) 
due to their proximity to the Galactic center ($R_{\rm{GC}} < 3$ kpc).
In particular, NGC\,6522 is a very interesting GC located in Baade's Window. 
Located at a Galactocentric distance $\lesssim$ 1 kpc,
  likely in the foreground with respect to the Galactic center, NGC\,6522
 was suggested to be among the oldest clusters in the MW
\citep{Barbuy+09}.

An important issue in the analysis of stellar populations in globular clusters
is the presence of multiple stellar populations
(e.g., \citealp{Milone+15,Milone+17,Nardiello+15a,Nardiello+15b,Piotto+15}).
In this paper we address this point with optical \textit{Hubble Space Telescope} (\textit{HST})
bands, and we present the first photometric evidence for multiple stellar populations in NGC\,6522.

Based on \textit{VLT}/FLAMES\footnote{Fibre Large Array Multi Element Spectrograph (FLAMES)
is the multi-object, intermediate- and high-resolution spectrograph of the
Very Large Telescope (VLT)}
optical high-resolution spectroscopy, \cite{Barbuy+09,Barbuy+14} and \cite{Ness+14} analyzed NGC\,6522
and derived its metallicity and chemical abundances.
With the same instrument, \cite{Villanova+17} performed a similar analysis for NGC\,6626
and interestingly found the presence of at least two stellar populations in the cluster.
Very recently, using near-infrared high-resolution spectroscopy from the APOGEE
(Apache Observatory Galactic Evolution Experiment; \citealp{Majewski+16}),
  \cite{Schiavon+17} discovered one star in NGC\,6522 with a very high nitrogen abundance ([N/Fe]=$+1.04$),
  suggesting that this cluster hosts multiple stellar populations.
  An additional three stars analyzed by \cite{Fernandez-Trincado+18} were found with similar chemical abundances.
  Furthermore, \cite{Recio-Blanco+17} inferred a high Al abundance together with a low Mg abundance for one star,
  corroborating the hypothesis of more than one stellar population hosted by NGC\,6522.

Despite the accurate metallicity and chemical abundance determinations 
for NGC\,6522 and NGC\,6626, to date there are no analogous results for self-consistent
age, distance, and reddening determinations for these clusters.
The main reason is the lack of deep, proper-motion-cleaned color-magnitude diagrams (CMDs),
based on high-photometric-precision data with which to perform a statistical analysis.
Currently only the \textit{HST} 
(e.g., \citealp{Ortolani+07,Ferraro+16,Piotto+15})
and 8m class telescopes with multi-conjugated adaptive optical (MCAO) systems
(e.g., \citealp{Ferraro+09,Saracino+15,Saracino+16})
are able to collect images with such high quality for GCs,
spatially resolving stars with an FHWM $\lesssim$ 0.10 arcsec in their cores.
The primary goal of this work is to derive such physical parameters (age, distance, and reddening )
for NGC\,6522 and NGC\,6626
using a statistical isochrone fitting on accurate, proper-motion-cleaned \textit{HST}-based CMD.

These two clusters were not included in the
ACS Survey of GCs \citep{Sarajedini+07} or in the
HST UV Legacy Survey of Galactic GCs \citep{Piotto+15},
and consequently they were not listed in the most recent and comprehensive papers 
concerning dating Galactic GCs
\citep{DeAngeli+05,Marin-Franch+09,Dotter+10,VandenBerg+13,Wagner-Kaiser+16,Wagner-Kaiser+17}.
Although they are included in the large compilation of age, metallicity, and abundance pattern for
41 Galactic GCs presented by \citet{Roediger+14}, their ages
(15.0 $\pm$ 1.10 for NGC\,6522 and 14.0 $\pm$ 1.1 for NGC\,6626)
are marked as ``less certain'' 
values\footnote{These age values are not even cited in their online table -- 
  http://www.astro.queensu.ca/people/Stephane\_Courteau/roediger\\
  2013/index.html.}.

The only two age determinations for NGC\,6522 based on \textit{HST} data were carried out
by \citet{Meissner+Weiss06} and \citet{Barbuy+09}, in both cases using
the same \textit{HST}/WFPC2\footnote{Wide Field and Planetary Camera 2 (WFPC2).}
data in F439W and F555W bands collected in a snapshot program
\citep{Piotto+02}.
Since the CMDs generated with these data barely reach 1 mag below the
main-sequence turnoff (MSTO), the most solid results in these papers
were based on the magnitude difference between the MSTO and the
horizontal branch (HB) ($\Delta V^{\rm{HB}}_{\rm{TO}}$).
In fact, NGC\,6522 presented a remarkably high $\Delta V^{\rm{HB}}_{\rm{TO}}$ value,
suggesting that it is $\sim$ 2 Gyr older than 47 Tuc and M5 (NGC\,5904) when
the mean loci of these clusters are overplotted \citep{Barbuy+09}. 
Using the same stellar evolutionary models
(A Bag of Stellar Tracks and Isochrones -- BaSTI, \citealp{Pietrinferni+04}),
  \citet{Meissner+Weiss06} recovered an age between 13.9 and 16.1 Gyr,
whereas \citet{Barbuy+09} estimated $\sim$ 16 Gyr.

As concerns NGC\,6626, \citet{Testa+01} presented a relative age determination based on
\textit{HST}/WFPC2 data.
These authors analyzed the F555W, F555W--F814W CMD for
this cluster in comparison with three halo GCs with BHB (NGC\,2298, NGC\,5897, NGC\,6535)
whose metallicities are [Fe/H]$\sim -1.73$, $-1.70$, and $-1.83$, respectively, i.e.
about 0.4--0.5 dex lower than the value for NGC\,6626 ([Fe/H]$\sim -1.33$).
Measuring the color difference between the MSTO and an arbitrary point in the red
giant branch (RGB), they identified NGC\,6626 as the oldest cluster in their sample, 
1.2 $\pm$ 0.9 Gyr older than NGC\,2298 and 2.4 $\pm$ 1.6 Gyr older than NGC\,6535.
Assuming an age of $\sim$ 13.0 Gyr for these two clusters as determined by \citet{Dotter+10},
NGC\,6626 could be older than $\sim$ 14.0 Gyr. 

\begin{figure*}
  \centering
\includegraphics[angle=0,scale=0.22]{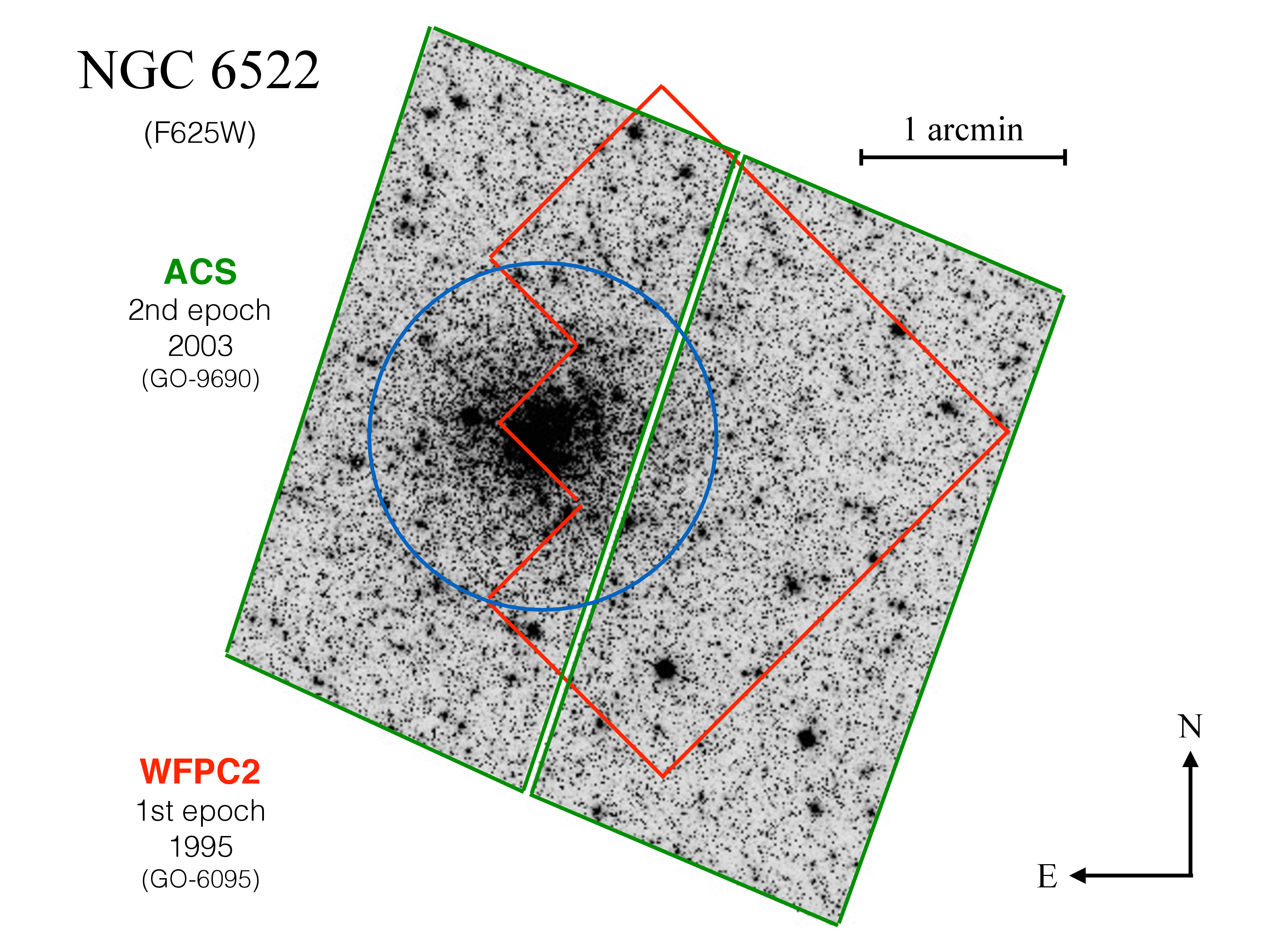}
\includegraphics[angle=0,scale=0.22]{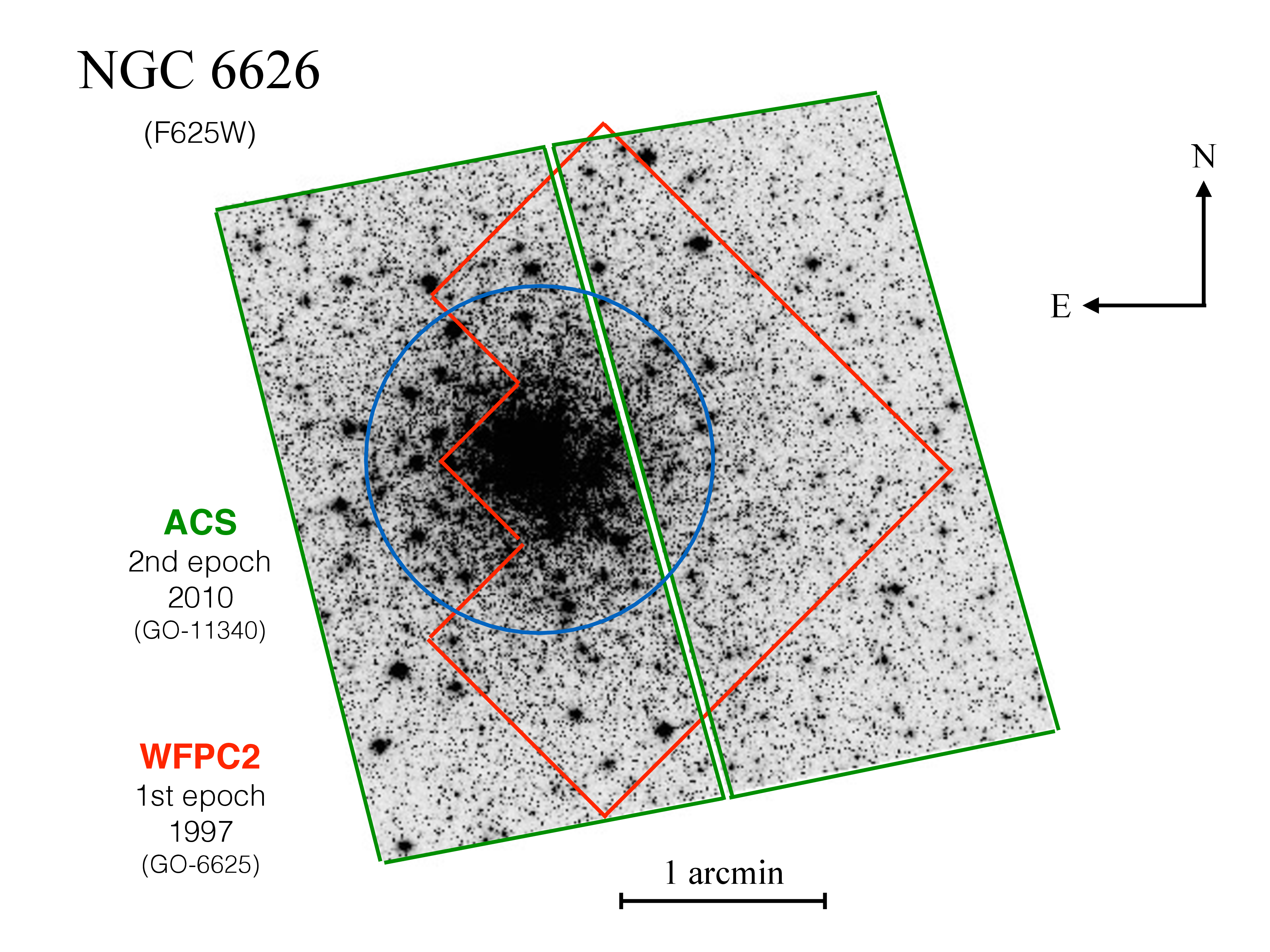}
\includegraphics[angle=0,scale=0.22]{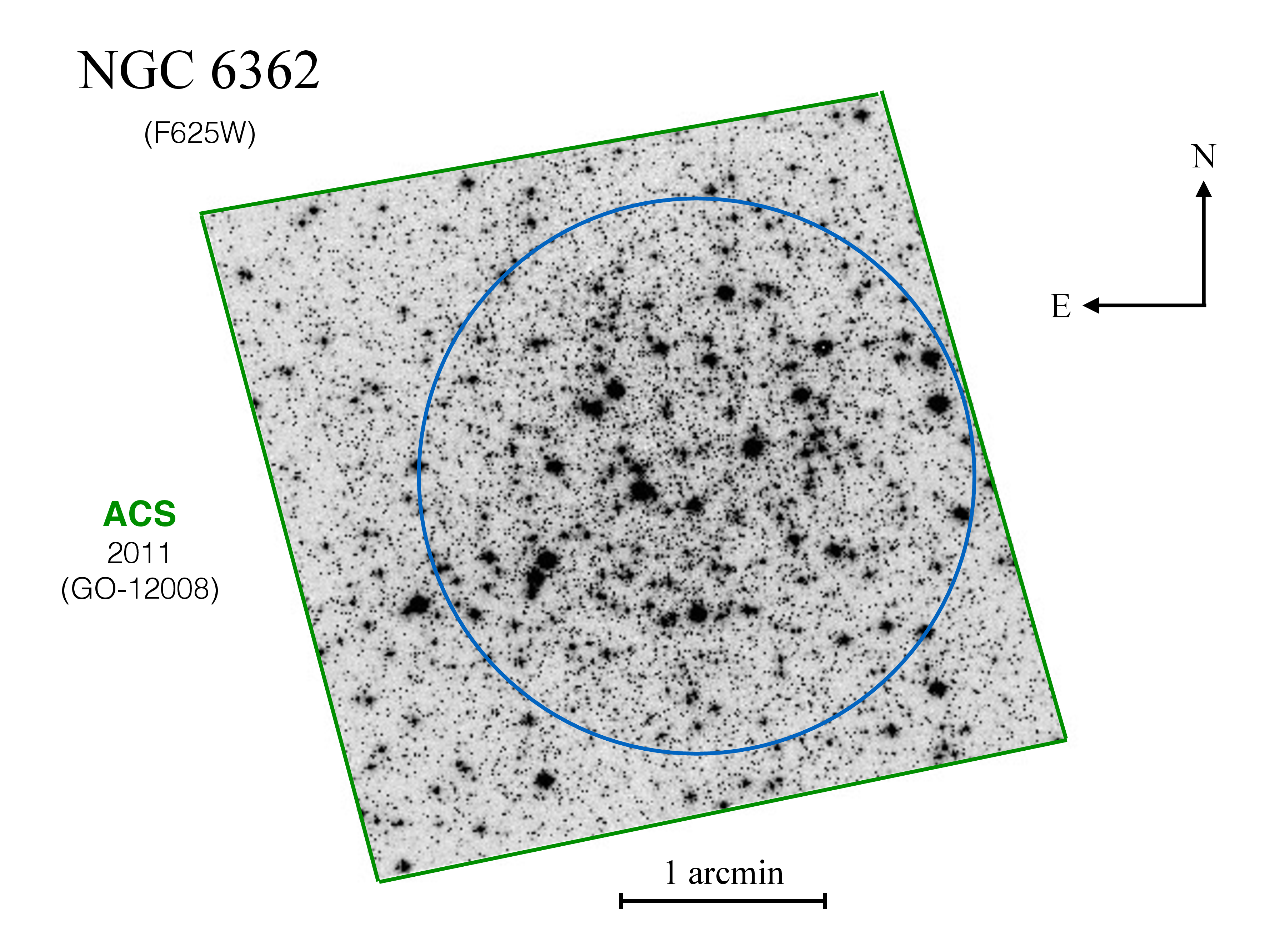}
\caption{\textit{HST} images of the GCs NGC\,6522 (left), NGC\,6626 (right), and NGC\,6362 (middle).
  All these images were collected with ACS/WFC in the F625W filter. North is up, east is left.
  The ACS/WFC field of view is 202 x 202 arcsec$^2$.
  The red contours correspond to the WFPC2 field of view of the first epoch for NGC\,6522 and NGC\,6626,
    whereas the blue circles are the adopted limiting radius defining the bulk of the cluster's stars.
}
\centering
\label{HST_images}
\end{figure*}

In order to determine some of the main physical parameters of NGC\,6522 and NGC\,6626 in a comparative way,
we retrieved \textit{HST}/ACS\footnote{Advanced Camera for Surveys (ACS).}
archive images in the filters F435W and F625W for these clusters.
Additional \textit{HST}/WFPC2 archive images for each cluster allowed us to generate
proper-motion-cleaned CMDs, drastically minimizing the effect of the field stellar contamination in our results.
The main method employed in the present work is based on statistical comparisons between observed
and synthetic fiducial lines, providing ages, distance moduli, and reddening values  self-consistently.
We used $\alpha$-enhanced BaSTI and Dartmouth isochrones with canonical ($Y \sim$ 0.25) and enhanced
helium abundance ($Y = 0.30$ for BaSTI; $Y = 0.33$ for Dartmouth) to generate
the synthetic fiducial lines.
The MSTO magnitude and the HB level are independently determined in order to give additional tests to
better constrain the cluster physical parameters. 
In particular, the RR Lyrae stars presented in the compilation by \citet{Clement+01}
(hereafter referred to as Clement's catalog -- 2017 edition\footnote{http://vizier.cfa.harvard.edu/viz-bin/VizieR?-source=V/150})
and in the OGLE catalog\footnote{http://ogledb.astrouw.edu.pl/~ogle/CVS/} (\citealp{Soszynski+14})
were used to determine the HB magnitude level.
Further detailed studies of the HB morphology in these clusters will be addressed in a future work.

For comparison purposes, we included the analysis of NGC\,6362,
a GC that has been extensively studied in the aforementioned \textit{HST} surveys and for which
there are \textit{HST}/ACS archive images in the F625W and F438W filters. 
It is an inner halo GC ($R_{\rm{GC}} = $ 5.1 kpc) located at a moderate Galactic latitude ($\sim$17\fdg6) and
therefore toward a region with low reddening ($E$(\textit{B}$-$\textit{V}) $\lesssim 0.10$) and low field stellar contamination.
Its metallicity was recently determined  by means of high-resolution spectroscopy
([Fe/H]$=-1.09 \pm 0.01$, \citealp{Mucciarelli+16}; [Fe/H]$=-1.07 \pm 0.01$, \citealp{Massari+17}),
confirming NGC\,6362 to be similar in metallicity to NGC\,6522 and NGC\,6626.
On the other hand, differently from NGC\,6522 and NGC\,6626, NGC\,6362 shows a red horizontal branch (RHB).

Furthermore, other works based on \textit{HST} data \citep{Dalessandro+14,Piotto+15,Milone+17}
and high-resolution spectroscopy \citep{Mucciarelli+16} demonstrated that NGC\,6362 hosts
multiple stellar populations.
Since NGC\,6362 is one of the analyzed clusters in \citet{DeAngeli+05}, \citet{Meissner+Weiss06},
\citet{Marin-Franch+09}, \citet{Dotter+10}, \citet{Paust+10}, \citet{VandenBerg+13}
 and \citet{Wagner-Kaiser+16,Wagner-Kaiser+17}, it is a key comparison object for the present paper.
 It was identified as one of the oldest Galactic GCs by
 \citet{Marin-Franch+09}, \citet{Paust+10}, \citet{VandenBerg+13} and \citet{Wagner-Kaiser+16,Wagner-Kaiser+17}.
  
The observations, proper-motion cleaning, and radial density profiles
are described in Section 2.
CMDs are presented in Section 3. In Section 4 the literature metallicities and abundances
are reported. In Section 5 the isochrone fitting method is presented,
and adopted stellar evolutionary models are briefly
described. In Section 6  cluster ages, distances, and reddening values are derived.
In Section 7 a discussion and a tentative identification of multiple
stellar populations in the sample clusters are presented. 
 Conclusions are drawn in  Section 8.

\section{Observations and Data Reduction}

\textit{HST} images of the bulge GCs NGC\,6522 and
NGC\,6626 and the inner halo cluster NGC\,6362 are employed in the present work (Figure \ref{HST_images}).
In order to build proper-motion-cleaned CMDs for the first two clusters,
images from two different epochs were retrieved from the \textit{HST}
archive.\footnote{https://archive.stsci.edu}

The \textit{HST} images of NGC\,6522 were collected as part of the program GO-9690
(PI: J. Grindlay), with the ACS/WFC.\footnote{Wide Field Channel (WFC) on the ACS.}
The observations of NGC\,6626 were performed using
the same camera during the program GO-11340 (PI: J. Grindlay).
For both clusters we reduced the data in F435W, F625W, and F658N filters.
For NGC\,6362 we employed data
in F438W obtained with the WFC3/UVIS\footnote{UVIS Channel on the Wide Field Camera 3 (WFC3).}
and in F625W (ACS/WFC) collected during 
GO-13297 (PI: G. Piotto) and GO-12008 (PI: A. Kong), respectively.

As a first step, all images were corrected for the imperfect charge transfer 
efficiency. 
The data reduction was carried out using the software described in
detail by \citet{Anderson+08}.
The point-spread function (PSF) model used in this work is a
perturbation of the library PSF by \citet{Anderson+King06};
in this way, we took into account the change of focus of the spacecraft.
We corrected the measured stellar positions for geometric distortion
by using the solution provided by \citet{Anderson+King06}.
We calibrated the magnitudes into the Vega-mag systems as in \citet{Bedin+05}.
Finally, we corrected the CMDs for differential reddening using the procedure described in detail by
\citet{Milone+12}.

An inspection of the theoretical magnitude predictions in the
filters F438W and F435W indicates a negligible difference of $\sim$0.01 mag. 
Even if small, this small difference was taken into account when the F435W magnitudes
of NGC\,6362 were converted to the F438W filter.

For NGC\,6522 we used as first-epoch images in F439W and F555W taken with the WFPC2
as part of the program GO-6095 (PI: S. Djorgovski) on 1995 September 9 ($t_{\rm I}=1995.69$).
As second epoch we used all the images in F435W, F625W, and F658N taken during GO-9690
(PI: J. Grindlay) on 2003 July 10 ($t_{\rm  II}=2003.52$).
The time baseline for the proper-motion measurements is $\sim 7.83$\,yr.
For NGC\,6626 we considered as first epoch the WFPC2 observations in
F555W and F814W carried out during  GO-6625 (PI: R. Buonanno) on 1997
September 12 ($t_{\rm I}=1995.70$), while as second epoch we used the
F435W and F625W observations taken for GO-11340 (PI: J. Grindlay) on
2010 April 26 ($t_{\rm II}=2010.32$). The time baseline is $\sim
14.62$\,yr.

\subsection{Proper Motions}
  
Stellar relative proper motions were measured for NGC\,6522 and NGC\,6266, 
in order to separate cluster
members and field stars. The approach used is the same as that
by, e.g.,
\citet{Anderson+06}, \citet{Bellini+10}, \citet{Libralato+14} and 
\citet{Nardiello+16}.

\begin{figure*}
\centering
\includegraphics[angle=0,scale=0.35]{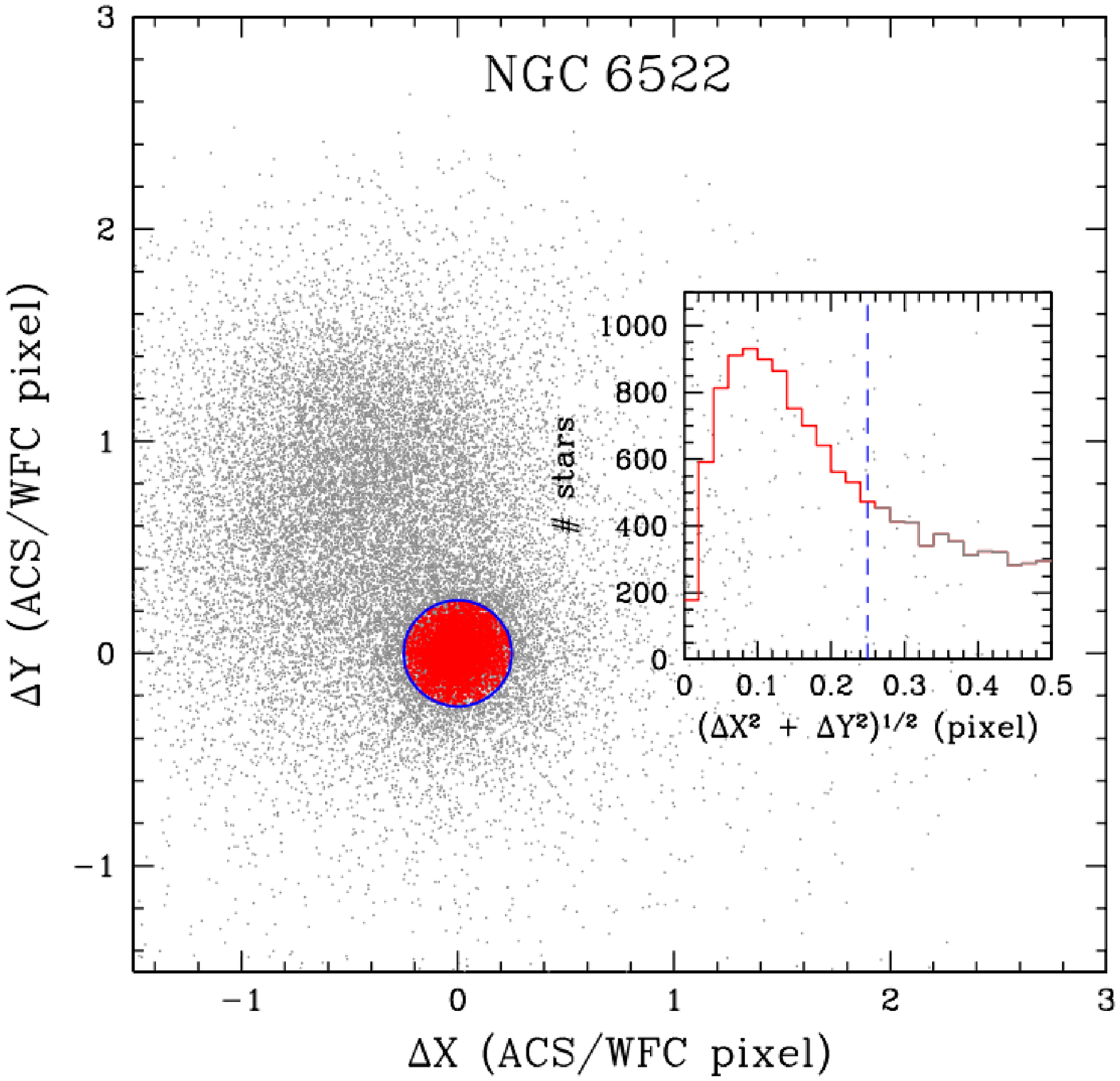}
\includegraphics[angle=0,scale=0.35]{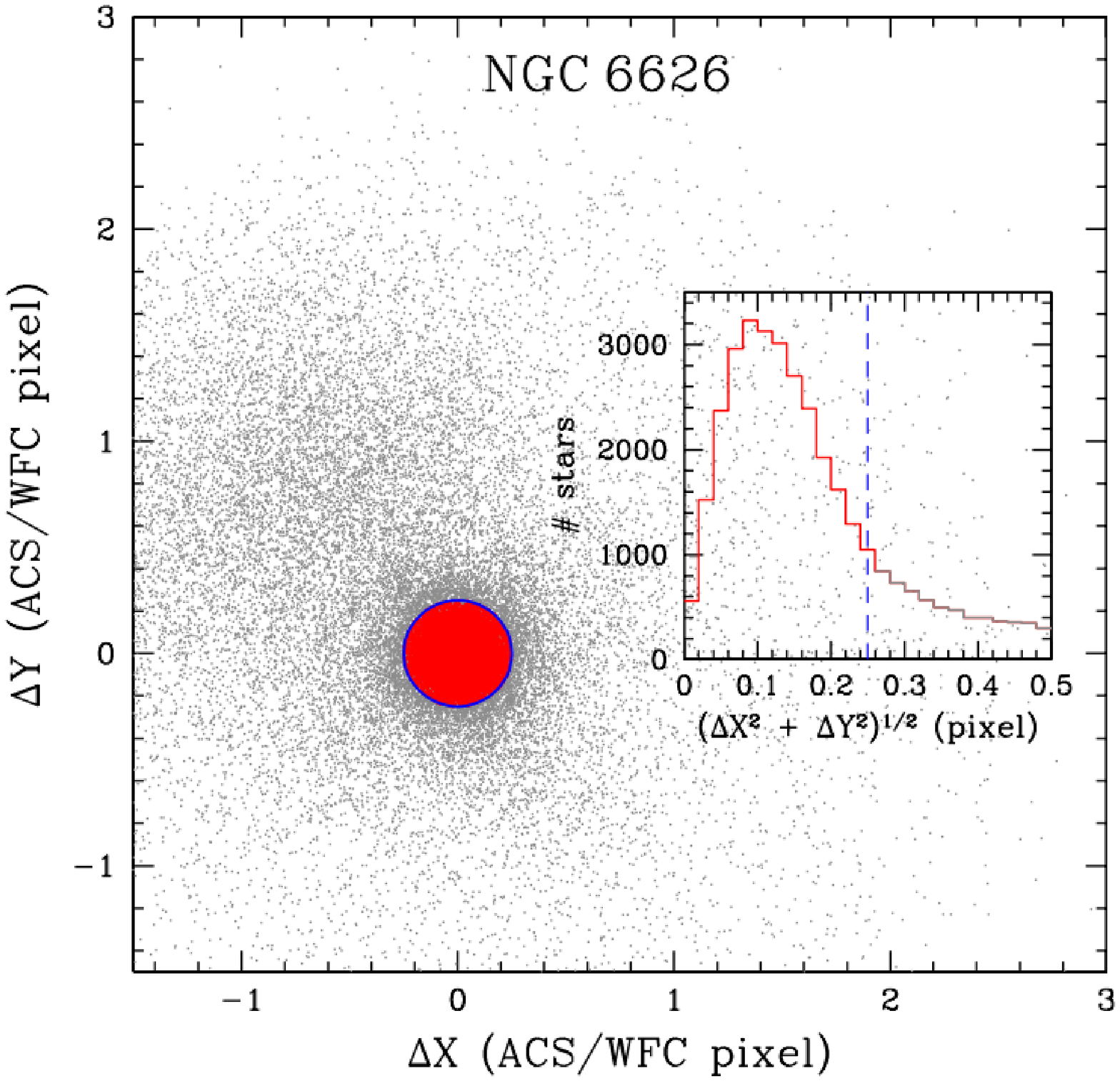}
\caption{
VPD of NGC\,6522 (left) and NGC\,6626 (right).
Likely cluster stars (red) and remaining stars (gray) are presented in each panel.
The histograms inserted in insets give
the number of stars as a function of $(\Delta X^{2}+\Delta Y^{2})^{1/2}$.
The vertical dashed line represents the cutoff limit to define the cluster members.
}
\centering
\label{VPD_pm}
\end{figure*}

Briefly, we used six-parameter, local linear transformations to transform stars' positions
as measured in the first-epoch reference frame into that of the second epoch and then computed 
the displacements. To compute the coefficients of these local transformations,
we considered a sample of close-by, likely cluster members
(e.g.,​ red giant branch [RGB], and subgiant branch [SGB] stars).
These stars were selected according to their
location in the CMDs and/or in the vector-point diagram (VPD).
​By using cluster stars, the stellar displacements in both coordinates ($\Delta X, \Delta Y$)
are computed relative to the cluster mean motion, and in the VPD the cluster distribution
is centered at (0,0), while the bulk of the field stars are located in a different region  (Figure \ref{VPD_pm}).
In this figure the number of stars as a function of total displacements $(\Delta X^{2}+\Delta Y^{2})^{1/2}$
is shown in the insets within each of the two main panels.
The cutoff in proper motion was chosen based on the visual inspection
of the final cleaned CMDs.

\begin{figure*}
\centering
\includegraphics[angle=0,scale=0.28]{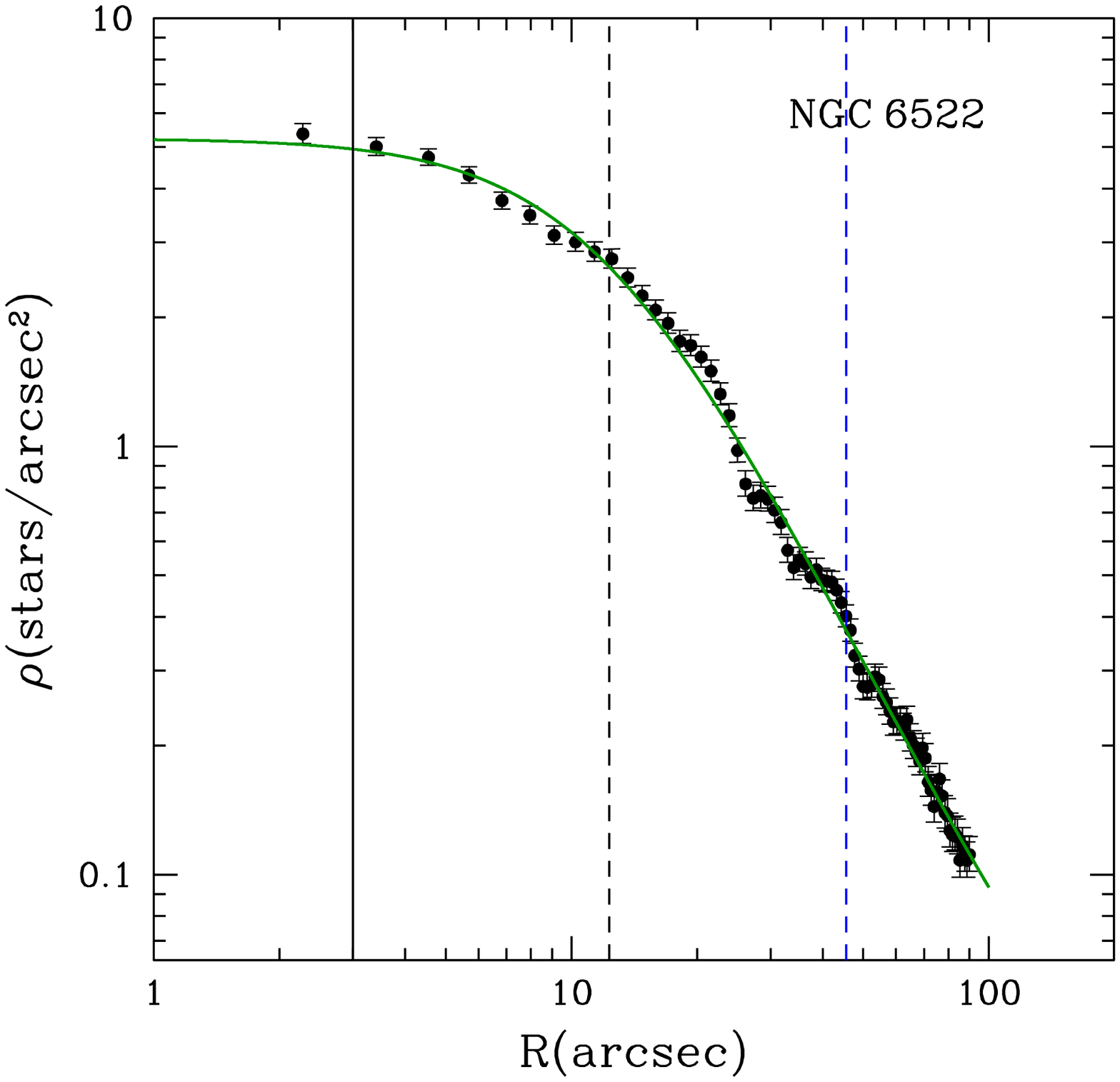}
\includegraphics[angle=0,scale=0.28]{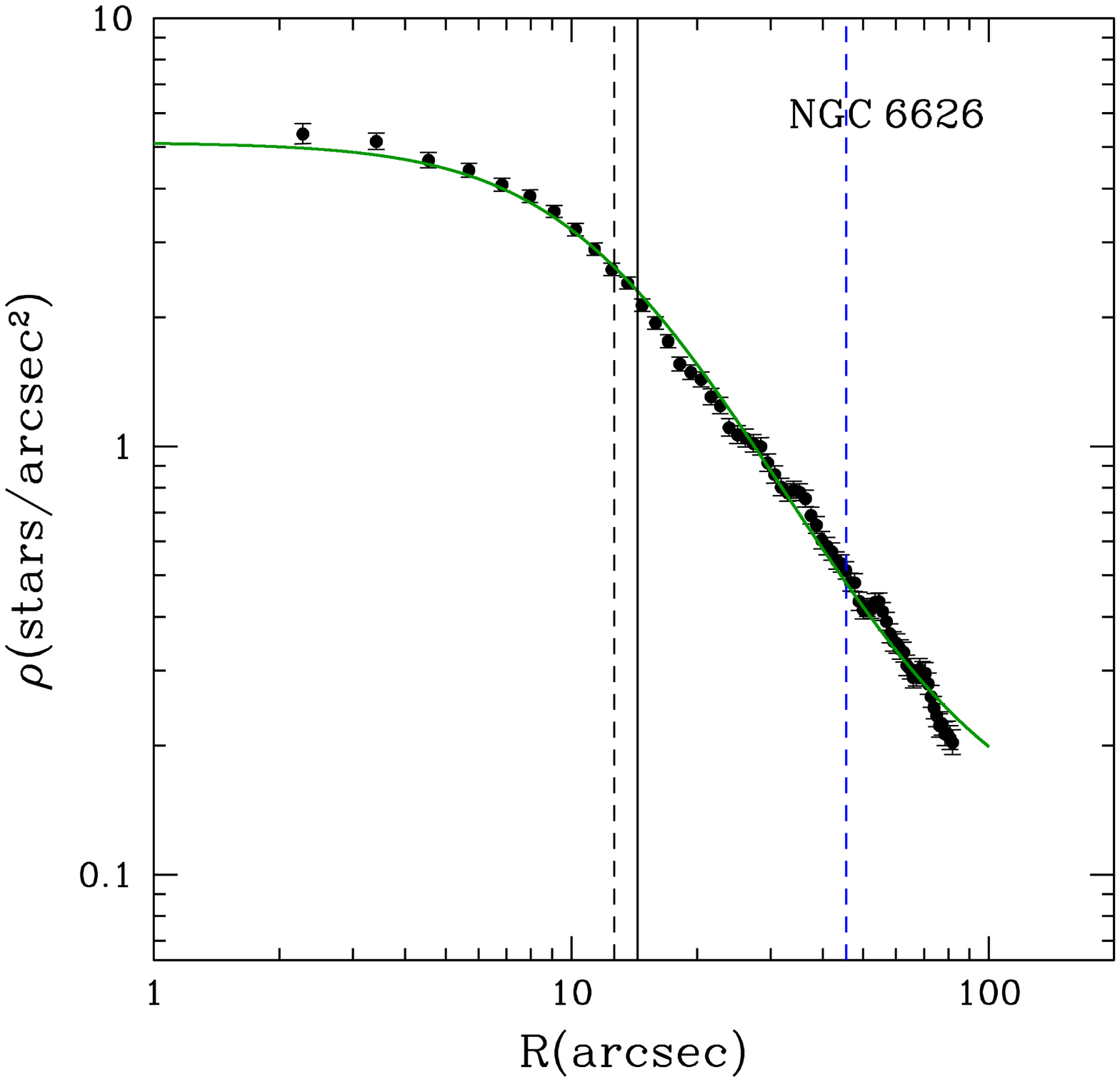}
\includegraphics[angle=0,scale=0.28]{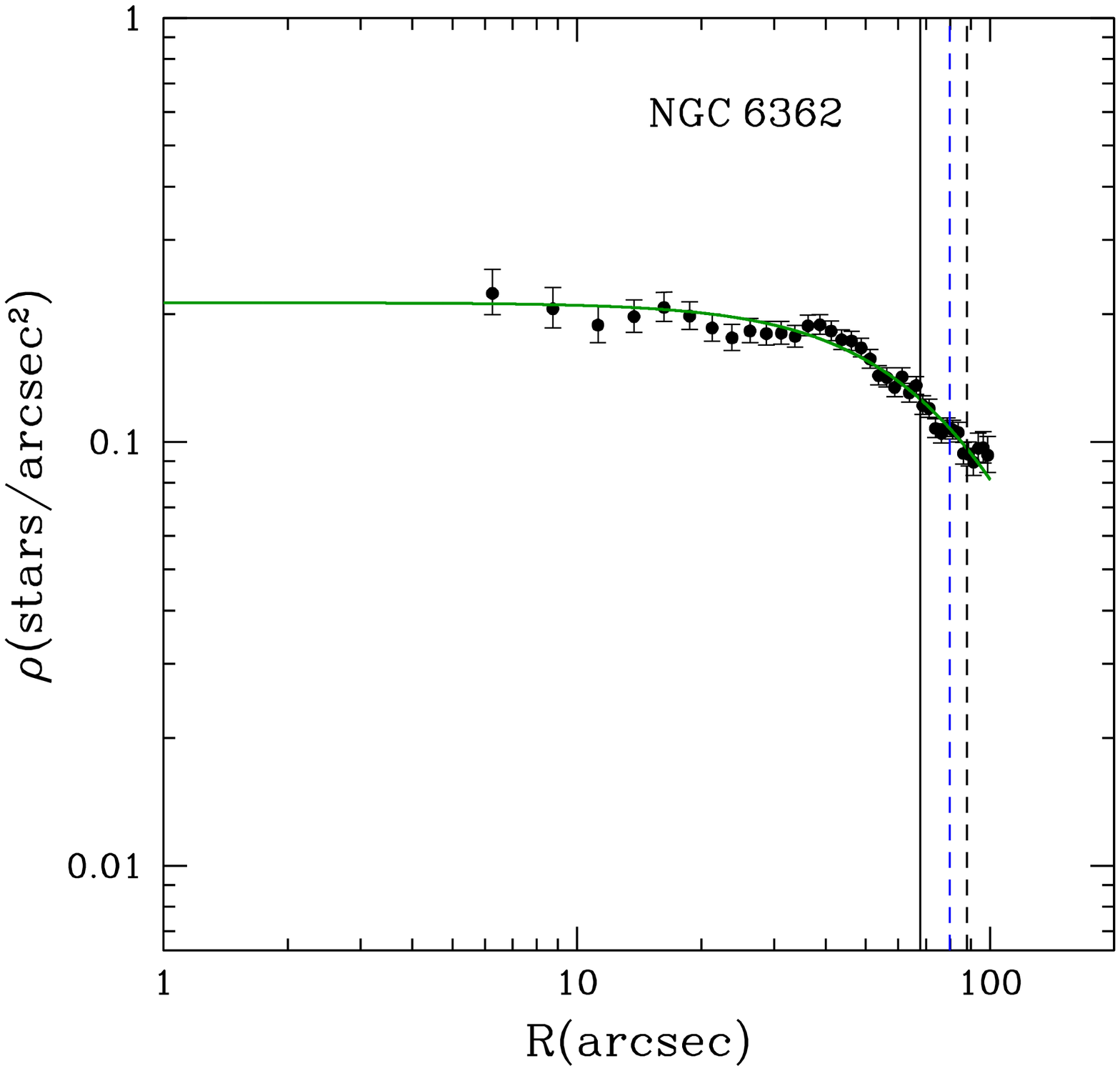}
\caption{Radial density profile of NGC\,6522 (left), NGC\,6626 (middle), and NGC\,6362 (right). 
The best fits for the King model are also presented (green lines). 
Vertical lines correspond to limiting radius as shown in Figure \ref{HST_images} (blue dashed line),
 core radius from King's model (black dashed line),
and core radius from \citet{Trager+95} (black solid line).} 
\centering
\label{RDPs}
\end{figure*}



\subsection{Radial Density Profiles}
\label{sect_RDPs}

The radial density profiles of NGC\,6522, NGC\,6626, and NGC\,6362
are presented in Figure \ref{RDPs}. 

They were determined by counting stars in a sequence of concentric rings using the F625W photometric catalog.
The center of the clusters was inferred by changing the center position until we found the ($X,Y$)
  coordinates where the density maximizes. 
The area of each ring was evaluated by means of a numerical approach, where a large number of artificial points
are generated in random positions.
Those points located inside the boundaries of the \textit{HST} instruments
(overlapping WFPC2 and ACS field of view for NGC\,6522 and NGC\,6626;
ACS for NGC\,6362) (see Figure \ref{HST_images}) are used to determine the area fraction covered
by them.
Only cluster stars brighter than F625W$_{\rm{MSTO}}$ + 1.0 mag were used, preventing incompleteness
effects.
In order to determine the core radius ($r_{\rm{c}}$) and the central star density ($\rho_{0}$), 
we adopted the two-parameter model from \citet{King62}:

$$ \rho(r) = \rho_{0} \frac{1}{1+(\frac{r}{r_{\rm{c}}})^{2}} .$$

A nonlinear minimum square fitting was employed to find the best solutions. 
The structural parameters from these fits are shown in Table \ref{tab_coord}.
NGC\,6522 and NGC\,6626 have very similar and small core radii ($\sim$ 12 arcsec),
as well as similar and high central star density ($\sim$ 5 stars arcsec$^{-2}$),
different from the ones presented for NGC\,6362
(91 $\pm$ 12 arcsec and 0.23 $\pm$ 0.03 stars arcsec$^{-2}$).
In Figure \ref{RDPs} we compare the core radius recovered by us and the ones
determined by \citet{Trager+95}. The agreement is very good for NGC\,6626 and NGC\,6362.
On the other hand, we obtained a core radius for NGC\,6522 significantly larger than the one
determined by \citet{Trager+95}.

We did not apply the three-parameter King's model
since the observed angular radii covered by the \textit{HST} images ($\sim 2$ arcmin)
are significantly smaller than the tidal radii of these clusters ($\gtrsim$ 5 arcmin).
In particular, only the core region of NGC\,6362 was analyzed because it
is a very extended cluster with a core radius $\gtrsim$ 60 arcsec.
For the two bulge clusters the analyzed area corresponds to about 10 times the core size.

\section{Color-Magnitude Diagrams}
\label{sect_CMDs}

\begin{figure*}
\centering
\includegraphics[angle=0,scale=0.65]{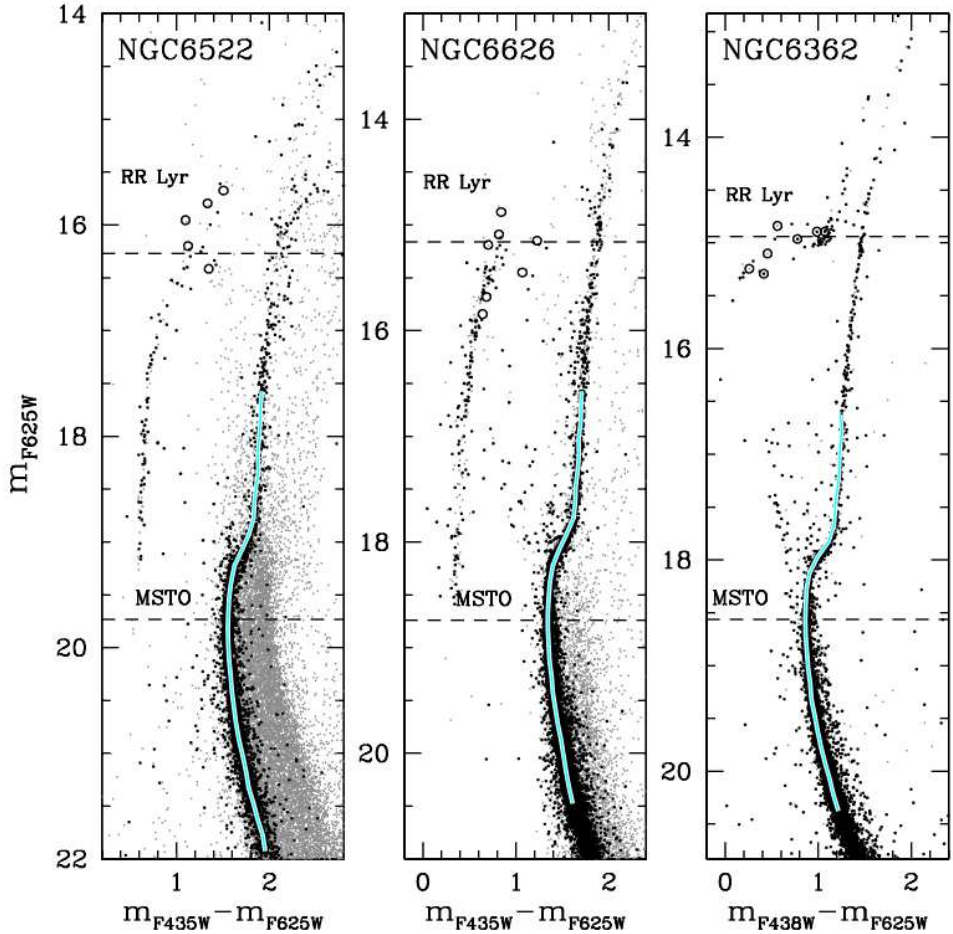}
\caption{F625W vs. F435W--F625W CMDs of NGC\,6522 and NGC\,6626, and F625W vs. F438W--F625W CMD of NGC\,6362.
  Cluster stars (black) and noncluster stars (gray) are shown in each panel.
  The fiducial lines (cyan solid lines) and the position of the RR Lyrae stars from
  Clement's catalog (open circles) that were identified in the \textit{HST}/ACS data
    are also presented.
  The horizontal dashed lines represent the MSTO magnitude (lower line)
  and  the average of the mean magnitudes for all RR Lyrae stars in Clement's catalog and the OGLE catalog (upper line)
  (See Section 3.2 for details).
}
\centering
\label{cmds}
\end{figure*}

Figure \ref{cmds} shows the CMDs of NGC\,6522, NGC\,6626, and NGC\,6362. 
In all CMDs, the main sequence (MS), SGB, and RGB are very well defined after field star 
decontamination, ranging along at least 7 mag.
Different magnitude and color limits are adopted for each cluster
 in order to place their MSTO level side by side, 
thus allowing direct comparisons.
An important difference between these clusters is related to the HB:
NGC\,6522 and NGC\,6626 have similar BHBs, 
whereas NGC\,6362 presents an RHB. 
Looking at Figure \ref{cmds}, it seems that the stars in the upper RGB are significantly
more scattered in the two bulge GCs than in NGC\,6362, indicating that there is a
possible saturation effect for stars brighter than the HB.
Blue stragglers seem to be present in the three clusters.

Figure \ref{cmds} also presents the fiducial line for each cluster, defined by
 the color median throughout the MS, SGB, and RGB, using magnitude bins of
 $\Delta$mag$ = 0.15$.

\subsection{MSTO}
The MSTO, defined as the bluest MS point, 
is the main CMD feature to characterize ages of stellar clusters.
In this work the MSTO point was determined by adopting the following procedure.
A denser sequence of fiducial points was determined by
applying small magnitude shifts (0.01 mag) to the original magnitude bins (0.15 mag),
therefore better sampling the MSTO region and reducing
the uncertainties in the MSTO position.
 The final MSTO magnitude in the F625W filter was defined as the 
average over all fiducial points with a color difference $\leq$ 0.010 mag
($\sim$ random uncertainty in color for each point) relative to the bluest point,
with the uncertainty in this value being provided by the standard deviation over 
 these points ($\leq$ 0.08 mag for all clusters).
 These determinations are presented in Table \ref{tab_photom}. 

\subsection{HB Level and RR Lyrae Stars}

For BHB clusters the HB magnitude at the RR Lyrae position is uncertain, particularly at the TO color,
where the HB presents few or even no stars, as can be seen in Figure \ref{cmds}.
For this reason the use of RR Lyrae to determine the HB magnitude in BHB clusters is highly
recommended, sometimes mandatory \citep{Zoccali+99}.  

To provide a first guess about the location of the RR Lyrae stars in the \textit{HST} CMDs,
we cross-matched our \textit{HST} data with the RR Lyrae stars presented in Clement's catalog.
We identified five RR Lyrae in NGC\,6522 and seven in NGC\,6626 and NGC\,6362 (Figure \ref{cmds}).
Although most of these RR Lyrae stars were not present in our multi-epoch photometry,
preventing their membership evaluation, they appear to be cluster members owing to their
magnitudes, similar to the HB one.
As expected, significant magnitude and color spreads can be seen in the \textit{HST} photometry
for these short-period variable stars. This is an effect of instantaneous measurements
at random phase positions in RR Lyrae light curves, since they present amplitudes that
can reach $\sim$ 1.0 mag in optical bands and periods of $\sim 0.2-0.7$ days.

\begin{figure}
\centering
\includegraphics[angle=0,scale=0.42]{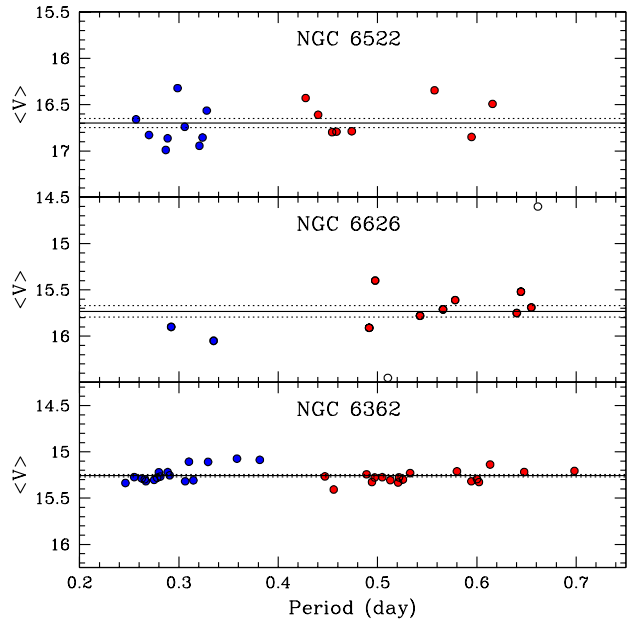}
\caption{ Mean $V$ magnitude vs. period of the RRab (red circles) and RRc (blue circles) Lyrae stars
  presented in NGC\,6522 (top panel),
  NGC\,6626 (middle panel), and NGC\,6362 (bottom panel).
  The mean of the mean $V$ magnitudes (solid line) and its standard deviation (dotted line)
  are shown in each panel. Two rejected field RR Lyrae stars toward NGC\,6626 (open circles) are also presented.
 }
\centering
\label{RRLyrae}
\end{figure}

In order to reduce the uncertainties in the HB level that would be introduced by determinations
using instantaneous magnitudes, we analyzed the $V$ mean magnitudes of the RR Lyrae stars presented
in Clement's catalog and the OGLE catalog (Figure \ref{RRLyrae}).

For NGC\,6522 there are 17 RR Lyrae stars in the OGLE catalog
(8 of Rab type and 9 of RRc type; \citealp{Soszynski+14}), and 10 of them are also present
in the compilation given in \citet[edition 2017]{Clement+01}.\footnote{In Clement's catalog the RRab and RRc types are designed as RR0 and RR1, respectively, as suggested by N.~N. Samus at the IAU XXVIth General Assembly, Prague 2006.}
For NGC\,6626 Clement's catalog presents only 10 RR Lyrae stars with mean magnitude
values in the $V$ filter (8 RRab and 2 RRc - \citealp{Wehlau+Butterworth90}).
Two RR Lyrae stars toward this cluster were rejected owing to their discrepant magnitudes,
being probably foreground or background field stars.
A large number of 35 RR Lyrae stars (18 RRab and 17 RRc; \citealp{Olech+01}) are available for NGC\,6362.
The OGLE catalog does not provide any entry for RR Lyrae stars in the other two GCs in our sample.
The mean of the mean magnitudes is also shown in Table \ref{tab_photom} and Figure \ref{RRLyrae},
as well as the standard deviation in these values.

The mean $V$ magnitudes of the RR Lyrae stars were converted into the F625W filter,
  allowing a consistent comparison with the MSTO, as previously determined by us.
The transformation of $V$ magnitudes presented in Clement's and OGLE catalogs
 into the F625W band has been done in the following steps:
conversion of the $V$ into the $R$ band and then from  $R$
to F625W using the \citet{Sirianni+05}  ACS calibration. 
First, the given magnitudes have been corrected for color
excess $E$(\textit{V}$-$\textit{R}) starting with our color excess $E$(\textit{B}$-$\textit{V}) 
(See Section \ref{results}) and the conversions from \citet{Fitzpatrick99}. 
Subsequently we adopted the transformations from $V$ to $R$
using PARSEC isochrone \citep{Bressan+12} colors.\footnote{Available at
http://stev.oapd.inaf.it/cgi-bin/cmd.}
The $R$ magnitudes so derived are then converted into
F625W based on \citet{Sirianni+05}.
As a consistency check, this transformation has also been obtained
using BaSTI isochrone tables for HB stars with the same average temperature of the RR Lyrae,
comparing the $V$  magnitudes with the corresponding F625W magnitudes.
This conversion requires interpolation of the data,
but the results are basically identical to the previous procedure.

In a conservative approach, we can assume that the uncertainties related
to the transformation from $V$ to F625W are $\sim$ 0.08,
including the ones from the reddening ($\sim 0.04$),
and in the transformation from $V$ to $R$ ($\sim 0.07$).
These uncertainties combined with those in the $V$ filter provide the
final values in Table \ref{tab_photom}.

Finally, we also calculated the magnitude difference between the MSTO and the HB
as determined by the RR Lyrae stars ($\Delta V^{\rm{HB}}_{\rm{TO}} $).
This is a very commom parameter used to constrain the ages of GCs
(e.g., \citealp{DeAngeli+05,VandenBerg+13}, and references therein) since
it is distance and reddening independent.
The $\Delta V^{\rm{HB}}_{\rm{TO}} $ parameter is particularly useful to determine relative ages
when the clusters have similar metallicities and are homogeneously analyzed in terms
of filters and method, as in the present case.

\begin{table*}
\begin{center}
\caption{Coordinates and Present Structural Parameters}
\begin{tabular}{lcccccc}
\tableline
\tableline
Cluster  & R.A. (J2000) & Decl. (J2000) & $l$ & $b$ & $\rho_{0}$ & $r_{\rm{c}}$ \\ 
         & (h m s) & ($\degr$ $\arcmin$ $\arcsec$) & ($\degr$) & ($\degr$) & (stars/arcsec$^2$) & (arcsec)  \\
\tableline
NGC\,6522 & 18:03:34.08 & -30:02:02.3 & -1.02 & -3.93 & 5.22 $\pm$ 0.13 &  12.32 $\pm$ 0.26 \\
NGC\,6626 (M28) & 18:24:32.89 & -24:52:11.4 & 7.80 & -5.58 & 5.01 $\pm$ 0.14 & 12.67 $\pm$ 0.31 \\
NGC\,6362 & 17:31:54.99 & -67:02:54.0 & 325.55 & -17.57 & 0.233 $\pm$ 0.026 & 91.2 $\pm$ 11.9 \\
\tableline 
\end{tabular}
\end{center}
\label{tab_coord}
\tablecomments{F625W magnitude cutoff for the King's model: F625W$_{\rm{MSTO}}$ + 1.0 (see Table \ref{tab_photom})}
\end{table*}

\begin{table*}
\begin{center}
\caption{Photometric Parameters from \textit{HST} Data and RR Lyrae Stars}
\begin{tabular}{lccccc}
\tableline
\tableline
Cluster  & MSTO    & RR Lyrae & RR Lyrae & RR Lyrae & $\Delta V^{\rm{HB}}_{\rm{TO}}$ \\
         & F625W & No. of Stars & $\langle V \rangle$ & $\langle$F625W$\rangle$ & $\langle$F625W$\rangle$ \\
\tableline
NGC\,6522 & 19.73 $\pm$ 0.05 & 17 & 16.70 $\pm$ 0.05 & 16.27 $\pm$ 0.09 & 3.46 $\pm$ 0.10 \\
NGC\,6626 (M\,28) & 18.74 $\pm$ 0.08 & 10 & 15.73 $\pm$ 0.06 & 15.16 $\pm$ 0.10 & 3.58 $\pm$ 0.13 \\
NGC\,6362 & 18.56 $\pm$ 0.08 & 35 & 15.26 $\pm$ 0.01 & 14.94 $\pm$ 0.08 & 3.62 $\pm$ 0.11 \\
\tableline 
\end{tabular}
\end{center}
\label{tab_photom}
\end{table*}

\begin{table*}
\begin{center}
\caption{Metallicity and Chemical Abundances from High-resolution Spectroscopy}
\begin{tabular}{lccccccc}
\tableline
\tableline
Cluster  & [Fe/H] & [O/Fe] & [Mg/Fe] & [Si/Fe]& [Na/Fe] & [$\alpha$/Fe] & References  \\ 
\tableline
NGC\,6522 & $-1.00~\pm$ 0.20 & +0.40       & +0.03       & +0.23 & +0.25     & +0.25 & B09 \\
          & $-1.15~\pm$ 0.15 & ...         & +0.43(2G)   & +0.42 & +0.28     & +0.36 & N14 \\
          & $-1.15~\pm$ 0.15 & ...         & $-0.21$(1G) & +0.42 & +0.28     & +0.36 & N14 \\
          & $-0.95~\pm$ 0.15 & +0.36       & $-0.07$     & +0.23 & +0.13     & +0.18 & B14 \\
          & $-1.06~\pm$0.07  & +0.33       &  +0.07      & +0.38 & ...       & +0.26 & F18 \\
\tableline
NGC\,6626 & $-1.29~\pm$ 0.01 & $-$0.36(2G) & +0.46(2G)   & +0.34 & +0.46(2G) & +0.37 & V17 \\
(M28)	  & $-1.29~\pm$ 0.01 & +0.27(1G)   & $-0.04$(1G) & +0.34 & -0.04(1G) & +0.38 & V17 \\
\tableline
NGC\,6362 & $-1.09~\pm$ 0.01 & ...         & ...         & ...   & +0.00(1G) & ...   & M16 \\
          & $-1.09~\pm$ 0.01 & ...         & ...         & ...   & +0.33(2G) & ...   & M16 \\
          & $-1.07~\pm$ 0.01 & ...         & +0.54       & +0.45 &  ...      & +0.32 & M17 \\
\tableline 
\end{tabular}
\end{center}
\label{tab_spec}
\tablecomments{B09 - \citet{Barbuy+09}; N14 - \citet{Ness+14}; B14 - \citet{Barbuy+14};
  F18 - \citet{Fernandez-Trincado+18};
  V17 -  \citet{Villanova+17}; M16 - \citet{Mucciarelli+16};
  M17 - \citet{Massari+17}
}
\end{table*}

\section{Physical Parameters from the literature}

\begin{table*}
\begin{center}
\caption{Age, Distances, Reddening, and Structural Parameters from Literature}
\begin{tabular}{lccccc@{}c@{}c@{}c@{}c@{}c@{}c@{}c@{}c@{}}
\tableline
\tableline
Cluster  & Age    & Refs & ($m-M$)$_{0}$ & $d_{\odot}$ & $R_{\rm{GC}}$  & ~~$E$(\textit{B}$-$\textit{V})~~ & $r_{\rm{c}}$ & c & $M_{\rm{V}}$ & Refs & Mass & Refs  \\
         &  (Gyr) &     &            & (kpc)  &  (kpc) &   & (arcsec) & ~~(log($\frac{r_{\rm{t}}}{r_{\rm{c}}}$))~~  &  &  & ~($\times 10^{4}$ M$_{\odot}$)~ & \\
\tableline
NGC\,6522 & $\gtrsim$ 14.0 & B09 & 14.43 & 7.7 & 0.6 & 0.48 & 3.1 & 2.50c & -7.65 & H96 & 5.93 & GO97 \\
& 15.0 $\pm$ 1.1 & MW06 (a) & 14.34 & 7.4 & 0.8 & 0.66 &  &  &  &  VFO10 &  &  \\
          & 13.8 $\pm$ 1.6 & MW06 (b) & 14.30 & 7.2 & $<$ 1 & 0.46 &  &  &  & T98 &  & \\
          &  &  & 14.52  & 7.8 & 0.6  & 0.48 &  &  & -7.67 & P02 &  &  & \\
\tableline 
NGC\,6626 & 14.2 $\pm$ 0.9  & T01, D10 &  13.70 & 5.5 & 2.7 & 0.42 & 14.4 & 1.67 & -8.16 & H96 & 44.2 & GO97 \\
(M28)     & $\sim$ 13.0 & V17 &   &  &  &  &  &  &  &  &  &  \\
\tableline 
NGC\,6362 & 12.5 $\pm$ 0.25 & VdB13 & 14.40 & 7.6 & 5.1 & 0.09 & 79.4 & 1.10 & -6.95 & H96 & 11.7 & GO97 \\
          & 12.5 $\pm$ 0.50 & D10 & 14.55 & 8.1 &  & 0.07 &  &  & -7.06 & D10 &  &  \\
          & 13.6 $\pm$ 0.6 & MF09 &  &  &  &  &  &  &  &  &  &  \\
          & 14.0 & P10 & 14.39 & 7.6 & 5.1 & 0.09 & 72 & 1.17 & -6.94 & P10 &  &  \\
          & 12.82$^{+0.03}_{-0.04}$ & WK16 & 14.56 & 8.1 &  & 0.09 &  &  &  & WK16 &  &  \\
          & 13.497$^{+0.003}_{-0.011}$ & WK17 & 14.44 & 7.7 &  & 0.10 &  &  &  & WK17 &  &  \\
          & 10.5 & MW06 (a) & 14.54 & 8.1 & 5.3 & 0.08 &  &  & -7.06 & P02 &  &  \\
          & 9.0 $\pm$ 0.5 & MW06 (b) &  &  &  &  &  &  &  &  &  &  \\

\tableline
\end{tabular}
\end{center}
\label{tab_param_lit}
\tablecomments{
  B09 - \citet{Barbuy+09};
  D10 - \citet{Dotter+10};
  GO97 - \citet{Gnedin+Ostriker97};
  H96 - \citet[update in 2010]{Harris96};
  MF09 - \citet{Marin-Franch+09}, 1.06 $\pm$ 0.05 $\times$ 12.8 Gyr (mean reference age using Dartmouth isochrones);
  MW06 - \citet{Meissner+Weiss06}, using their own isochrones (a) and BaSTI (b) isochrones;
  P10 - \citet{Paust+10};
  P02 - \citet{Piotto+02};
  T98 - \citet{Terndrup+98};
  T01 -  \citet{Testa+01}, 1.2 $\pm$ 0.90 Gyr older than NGC\,2298, taken as 13.0 Gyr from D10;
  VFO10 - \citet{Valenti+10};
  VdB13 - \citet{VandenBerg+13};
  V10 - \citet{Villanova+17}, age comparable to M12 owing to their similar metallicity and BHB; 
  WK16 - \citet{Wagner-Kaiser+16}, for two populations with $Y=0.220$ (1G) and $0.265$ (2G);
  WK17 - \citet{Wagner-Kaiser+17}, for a single population with $Y=0.327$ and adopting [$\alpha$/Fe]=+0.40.
}
\end{table*}

\subsection{Metallicity and Alpha-elements}

Metallicities and main abundance ratios available are reported in Table \ref{tab_spec}.
A spectroscopic analysis of eight RGB stars in NGC\,6522 using
the FLAMES-GIRAFFE spectrograph, with  $R\sim$22,000, was presented in \citet{Barbuy+09}.
The authors found a metallicity of [Fe/H]$=-1.0 \pm 0.2$ together with the $\alpha$-element enhancements
(see Table \ref{tab_spec}).
Using the same FLAMES-GIRAFFE data, \cite{Ness+14} reanalyzed the eight stars and measured
[Fe/H]=$-1.15$, and a somewhat higher enhancement of $\alpha$-elements.
An improved analysis of four of the same
 RGB stars by \citet{Barbuy+14} was obtained using
the FLAMES-UVES spectrograph at $R\sim$45,000, and a higher signal-to-noise ratio. 
This resulted in [Fe/H]$=-0.95 \pm 0.15$, and $\alpha$-enhancements similar
to those given in \citet{Barbuy+09}.
Based on the DR13 release of APOGEE,
\citet{Fernandez-Trincado+18} find [Fe/H]$=-1.06 \pm 0.06$,
and [$\alpha$/Fe]=+0.26 (mean of O, Mg, Si) for second-generation stars.
As concerns NGC\,6626, \citet{Villanova+17} recently analyzed 21 stars observed with FLAMES-UVES.
They obtain [Fe/H]$=-1.29$, and abundances for 21 elements.

For NGC\,6362 the spectroscopic analysis carried out
by \citet{Mucciarelli+16} gives values of metallicity [Fe/H] and [Na/Fe]. 
Using sodium as a key indicator of stars from first and second generations
  (e.g., \citealp{Campbell+13}), \cite{Mucciarelli+16} and \cite{Villanova+17}
  provided two distinct [Na/Fe] values for NGC\,6362 and NGC\,6626, respectively.
For NGC\,6522 no Na excess is found in \citet{Barbuy+09, Barbuy+14},
except for star B-8 in \citet{Barbuy+09}. 
\citet{Ness+14} found three stars Na-normal and five stars with Na excess,
partly due to their lower metallicities relative to \citet{Barbuy+14}.
Note that for NGC\,6522 [$\alpha$/Fe]=[$\langle$(O$+$Mg$+$Si$+$Ca$+$Ti)$\rangle$/Fe], whereas for
 NGC\,6626 [$\alpha$/Fe]=[$\langle$Mg$+$Si$+$Ca$+$Ti$\rangle$/Fe], since oxygen abundances are not derived for the latter. 

 The [$\alpha$/Fe] values reported in Table \ref{tab_spec}
 justify the adopted [$\alpha$/Fe]$\sim+$0.4 available in the isochrone sets.
Such $\alpha$-enhancements are of the same order as measured in other clusters of the Galactic bulge
(e.g., Table 4 of \citealp{Bica+16}).
 
Finally, it must be pointed out that since Mg and Si are important
electron donors, they contribute to the formation of the H$^{-}$ ion, which is the
main opacity source at the temperatures of G--K type stars, therefore
affecting the effective temperature scale of
both MSTO and (mostly) RGB stellar models. The trend is that by
increasing the Mg and/or Si abundance at a given [Fe/H], the RGB becomes cooler: an
increase of 0.4 dex causes a decrease of about 100--150 K in $T_{\rm eff}$.
Because Mg and Si (together with O and Ne) affect the difference in $T_{\rm eff}$ (and
hence color) between the MSTO and the lower RGB at a fixed age, the use
of such a diagnostic to determine the relative ages of star clusters
having very similar [Fe/H] values will yield reliable results only if
the cluster-to-cluster differences in the abundances of these elements
are small, or if the effects of such differences are taken into account
(see \citealp{VandenBerg+12} for a detailed discussion on this issue).
As a warning, we note that -- as listed in Table \ref{tab_spec} -- the Mg and Si abundances
  in these GCs might be lower than that of halo clusters.

\subsection{Age, Distance, Reddening, and Structural Parameters}

In Table \ref{tab_param_lit} we compile literature results for the sample clusters,
including ages, solar and galactocentric distances, reddening, absolute magnitudes,
core radii, and concentration parameter.
NGC\,6522  and NGC\,6626 are within the bulge volume,
and  NGC\,6522 is located at $<$ 1 kpc from the Galactic center in Baade's Window. 

As for the ages, NGC\,6522 was found to be very old ($\gtrsim$ 13.8 Gyr) 
in two studies available
using \textit{HST} previous data \citep{Meissner+Weiss06,Barbuy+09}.
For NGC\,6626 there is only one relative age determination \citep{Testa+01},
suggesting that is 1.2 $\pm$ 0.9 Gyr older than NGC\,2298 (13.0 Gyr, as determined by \citealp{Dotter+10}).
Recently, \citet{Villanova+17} argued that this cluster should be as old as M15 ($\sim$ 13.0 Gyr)
owing to their similar BHB at the same metallicity. 
The inner halo cluster NGC\,6362 has been studied more extensively in the literature and can be used as a reference
with respect to other studies, and conversely as a comparison probe with the present clusters.
Most results for this cluster indicate an age $\gtrsim$ 12.5 Gyr, therefore
 classifying it as an old GC.
The only exception is the work from \citet{Meissner+Weiss06},
where a younger age ($\sim$ 10 Gyr) for NGC\,6362 is given,
 significantly younger (by $\sim 4$ Gyr) than NGC\,6522. 

The two bulge clusters have $E$(\textit{B}-\textit{V}) $>$ 0.40, while NGC\,6362 has a low
reddening. On the other hand, 
NGC\,6522 and NGC\,6362 present similar solar distances ($\sim$ 7--8 kpc),
$\sim 2$ kpc farther than NGC\,6626.

The structural parameters indicate that NGC\,6522 is a core-collapse
 cluster \citep{Terndrup+98}, and the most compact and least massive object
 ($\sim6\times10^{4}M_{\odot}$) in our sample  \citep{Gnedin+Ostriker97},
 whereas NGC\,6626 is the most massive ($\sim4\times10^{5}M_{\odot}$)
and NGC\,6362 is the least compact one.

\section{Isochrone fitting}
\label{isot_fits}

\subsection{Isochrone Models}
\label{isochronemodels}

In order to compare the \textit{HST} data for the sample clusters with theoretical models, we selected 
$\alpha$-enhanced isochrones with canonical ($Y\sim0.25$) and helium-enhanced ($Y=0.30-0.33$) isochrones from the 
Dartmouth Stellar Evolutionary Database (DSED; \citealp{Dotter+08}) and from 
BaSTI models \citep{Pietrinferni+06}.

\begin{figure*}
\centering
\includegraphics[angle=0,scale=0.42]{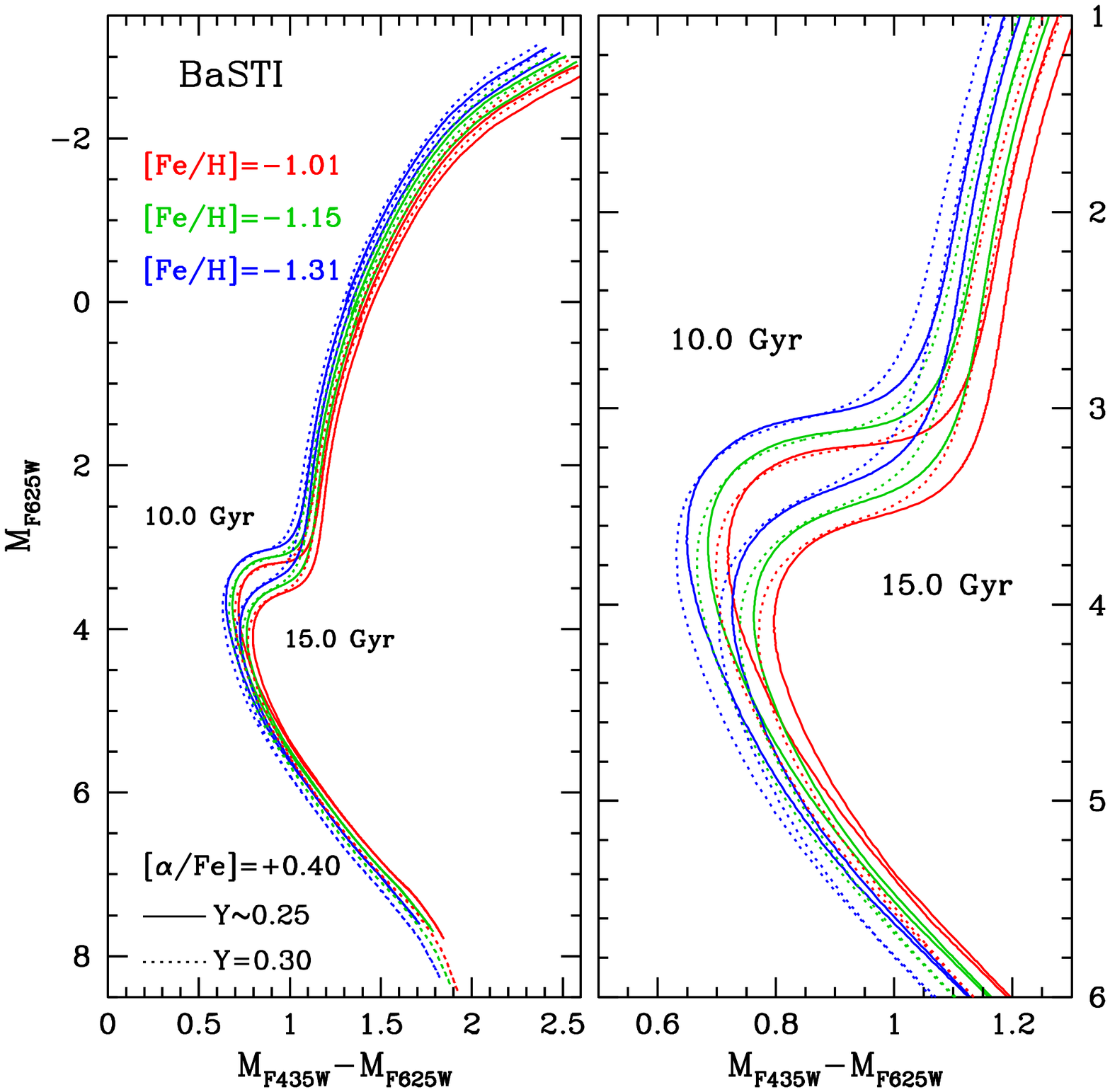}
\includegraphics[angle=0,scale=0.42]{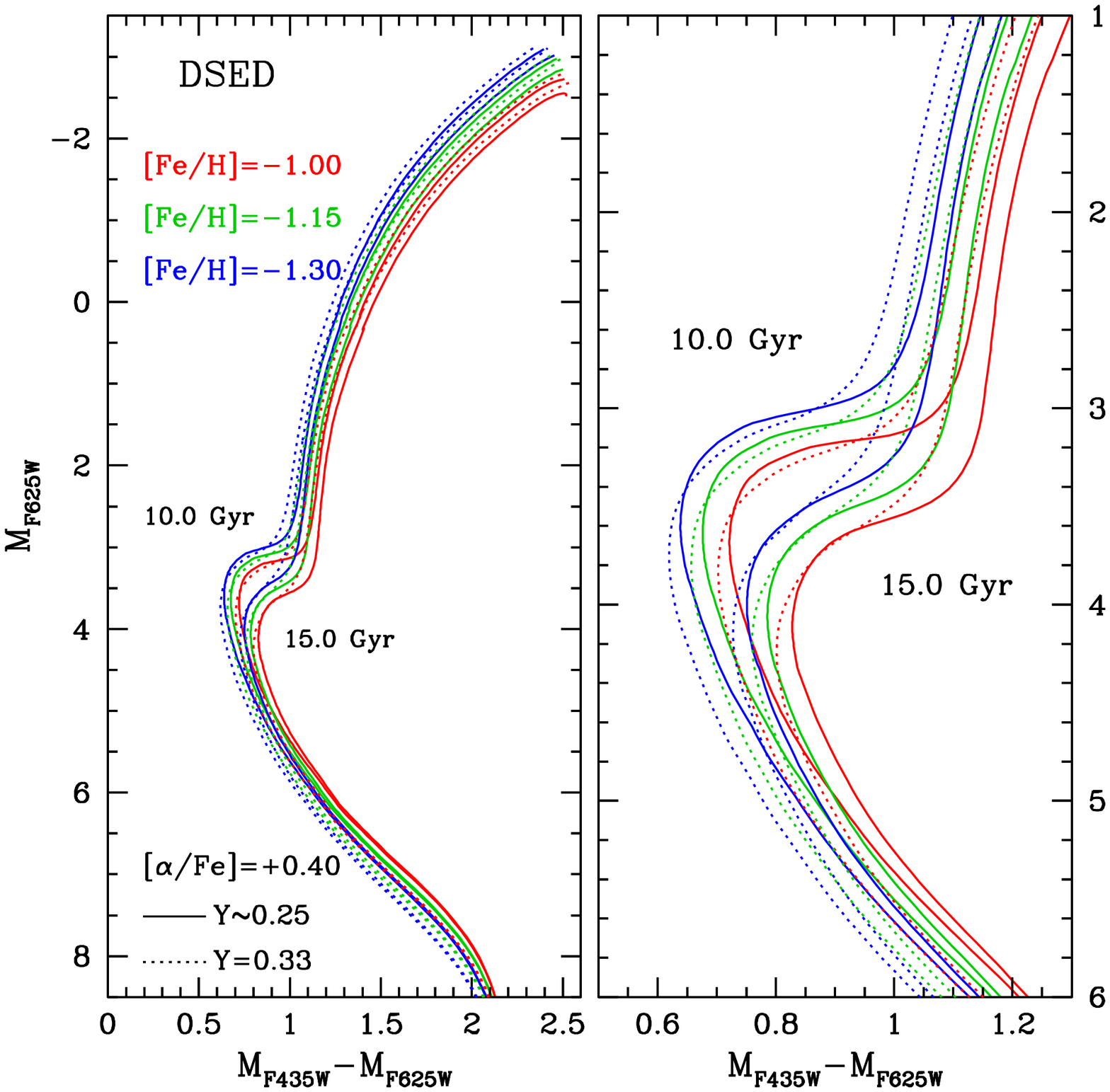}
\caption{Isochrones from BaSTI (left panels) and Dartmouth (right panels)
  showing the  metallicity effects for two ages (10.0 and 15.0 Gyr).
  All these isochrones are $\alpha$-enhanced ([$\alpha$/Fe]=+0.40]).
 Isochrones with [Fe/H]$\sim -$1.0 and canonical ($Y\sim 0.25$; solid lines) and enhanced 
   ($Y=0.30$ for BaSTI models, $Y=0.33$ for DSED models; dotted lines) helium abundances are presented.
 }
\centering
\label{isochrones}
\end{figure*}

Based on the spectroscopic results for these clusters, isochrones
 with [Fe/H]=$-1.0$, $-1.15$ and $-1.30$ and 
[$\alpha/$Fe]$=+0.40$
were employed, covering ages from 10.0 to 15.0 Gyr (in steps of 0.50 Gyr).
The effects of age, metallicity, and helium enhancement in DSED and BaSTI isochrones are 
illustrated in Figure \ref{isochrones}.
Essentially helium-enhanced isochrones predict bluer MS and RGB, as well as fainter MSTO, 
than the isochrones with standard helium abundance. 
A $\Delta Y \sim 0.05-0.08$ can produce an effect as large as $\sim 0.05$ in (F435W--F625W) color
for the MS or RGB stars, and about 0.20 mag (F625W) in the MSTO magnitude,
resembling a difference of $\sim$ 0.15 dex in [Fe/H] or $\sim$ 2.5 Gyr in age.
Therefore, it is expected that isochrone fits using helium-enhanced models will recover higher reddening and lower
distance values, as well as younger ages. 
Furthermore, helium enhancement produces a remarkable change in the SGB shape, which becomes
significantly steeper. 

 It is important to point out that the BaSTI models
 have been computed by neglecting the effect of
atomic diffusion, which is included in the DSED models
(when computing BaSTI models it was chosen not to include this
 effect owing to the uncertainty on the atomic diffusion coefficients).
This introduces an offset in the age scale based on the two distinct model sets:
the inclusion of the diffusion reduces the age by about 0.9 Gyr
at the metallicity  of the sample GCs
(\citealp{Cassisi+98,Cassisi+99} and references therein).
Therefore, the BaSTI-based ages should be reduced by 0.9 Gyr, and this completely
removes the apparent contradiction with the age of the universe
  (13.799 $\pm$ 0.021 Gyr; \citealp{Planck16}). 

\begin{figure}
\centering
\includegraphics[angle=0,scale=.40]{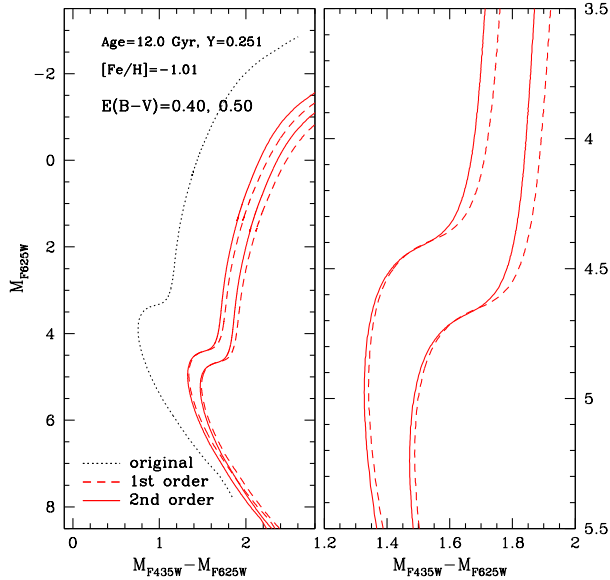}
\caption{CMD correction for the reddening effect using a uniform $E$(\textit{B}$-$\textit{V})
  (first order; dashed lines) and a temperature-dependent one (second order; solid lines).
  A single $\alpha$-enhanced BaSTI isochrone with 12 Gyr, $Y=0.251$ and [Fe/H]$=-1.01$ is converted
    from its reddening-free position (dotted line) to the ones corresponding to the reddening values
    $E$(\textit{B}$-$\textit{V})=0.40 and 0.50.
}
\label{ACS} 
\end{figure}

Before using theoretical isochrones to fit the cluster CMDs,
we had to consider carefully an important issue, related to the high extinction along the 
line of sight to some clusters in our sample, as in the case of NGC\,6522 and NGC\,6626, for which $A_{V} > 1$. 
It is well known that the ratio between the extinction in a given photometric band ($A_{\lambda}$) 
and $A_{V}$ depends on the flux distribution 
of the stellar source and is in principle dependent on parameters such as 
the effective temperature, surface gravity, and chemical composition (see, e.g., \citealp{Bedin+05};
\citealp{Ortolani+17}, for discussions and data specific for the ACS photometric filters).
As long as $A_{V}$ is small, this effect is negligible and a single value of
 $A_{\lambda}/A_{V}$ can be safely applied along the whole isochrone. 
 But the high value of the extinction in the case of NGC\,6522 and NGC\,6626  
 makes it necessary to consider, in the fits to the observed CMD, the variation 
 of $A_{\lambda}/A_{V}$ along the isochrones due to (mainly) the change in the stellar effective temperature. 
 In order to account for this effect when needed, we proceeded as follows: we first used the
 CMD 3.0 web interface,\footnote{http://stev.oapd.inaf.it/cgi-bin/cmd}
 which implements the results by \citet{Girardi+08}, 
 to determine the extinction in the relevant photometric filters, covering the full range of 
 $T_{\rm{eff}}$ values of our isochrones and zero-age horizontal branch (ZAHB),
 for a suitable value of the metallicity, and 
 for varying values of the extinction; these final extinction values in the F625W and F435W bands 
 were then applied to the theoretical  isochrones transformed to the ACS system.
These ${\rm T_{eff}}$-dependent extinction corrections have the effect of steepening
the RGB by $\Delta$(F435W--F625W) $\sim$ 0.05, as shown in Figure \ref{ACS}; indeed, this correction is 
very relevant for present work, because the effect on the RGB and SGB shape is quite similar
to that related to a variation in the initial He abundance.

\subsection{Statistical Fiducial Line Comparisons}
\label{fitting}

The age, reddening, and distance modulus of each cluster were derived from
statistical comparisons between synthetic fiducial lines
and the observed ones. 
Similar procedures have been applied to analyze \textit{HST} CMDs 
of Galactic GCs (e.g., \citealp{Marin-Franch+09,VandenBerg+13})
and populous LMC clusters (e.g., \citealp{Kerber+07}). 
The synthetic fiducial lines were determined by the color 
median positions at each magnitude bin
of synthetic CMDs encompassing the MS, SGB, and RGB. 
These CMDs were generated using DSED and BaSTI isochrones with 
[$\alpha$/Fe]=+0.40, [Fe/H]=$-1.0$ (for NGC\,6522, NGC\,6362) and $-1.15$ (for NGC\,6522, NGC\,6362),
and [Fe/H]=$-1.3$ (for NGC\,6626)
and different helium abundances (from $Y\sim0.25$ to 0.33).
The synthetic CMDs explored a wide and regular model grid, typically covering 
$\Delta$(age)=5.0 Gyr, $\Delta$($E$(\textit{B}$-$\textit{V}))=0.20, $\Delta$(($m-M$)$_{0}$)=0.40, 
in steps of 0.50 Gyr, 0.02 mag, and 0.02 mag respectively.
Furthermore, educated guesses for photometric uncertainties (similar to the ones determined for the bulk of stars), 
binarity (20\%), and initial mass function (Salpeter) were used in order to reproduce the
observed CMD features. 
The $\chi^{2}$ statistics was employed to compare the 
model (mod) and data (obs) colors, being computed for the $N_{\rm{bin}}$
magnitude bins along the fiducial line according to the expression

$$\chi^{2} = \frac{1}{N_{\rm{bin}}-1}\sum_{i=1}^{N_{\rm{bin}}}\left[\frac{(\rm{color})_{\rm{obs},\textit{i}}-(\rm{color})_{\rm{mod},\textit{i}}}{\sigma_{\rm{color}}}\right]^{2} ,$$

where $\sigma_{\rm{color}}$ is the dispersion in the median 
color position for the $i^{\rm{th}}$ magnitude bin in the model (typically $\sim$ 0.02). 

To determine the final values of the fit parameters, as well as to study the confidence intervals
and correlations between them, we applied the Markov chain Monte Carlo (MCMC) sampling technique.
For that purpose, we used the \texttt{emcee} code \citep{Foreman-Mackey+13} to sample the posterior
probability in the three-dimensional parameter space, assuming $\mathrm{likelihood}\propto \exp[-\chi^2/2]$
and a uniform prior probability within the acceptable physical ranges. 

\begin{figure*}
\centering
\includegraphics[angle=0,scale=0.40]{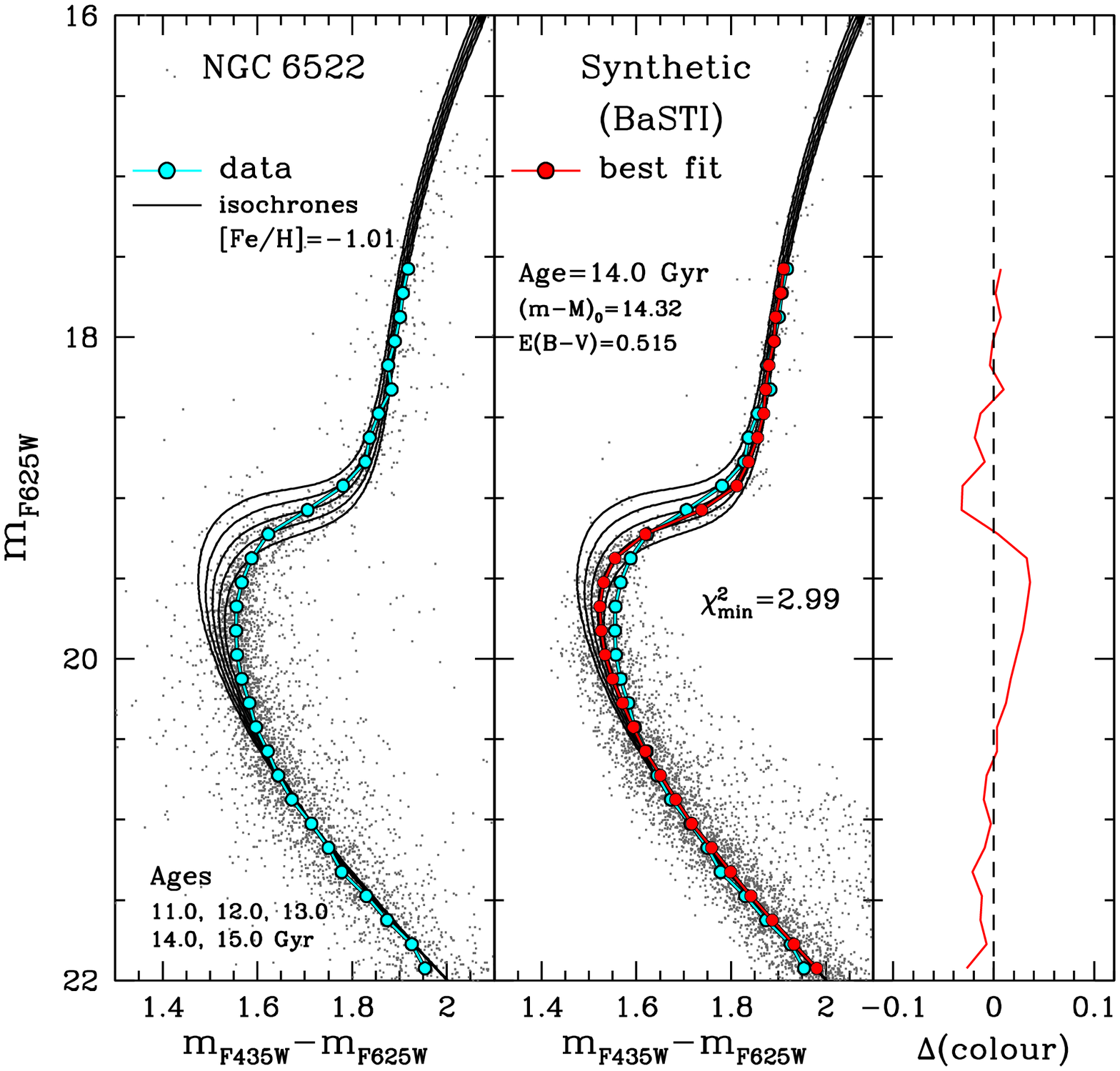}
\includegraphics[angle=0,scale=0.40]{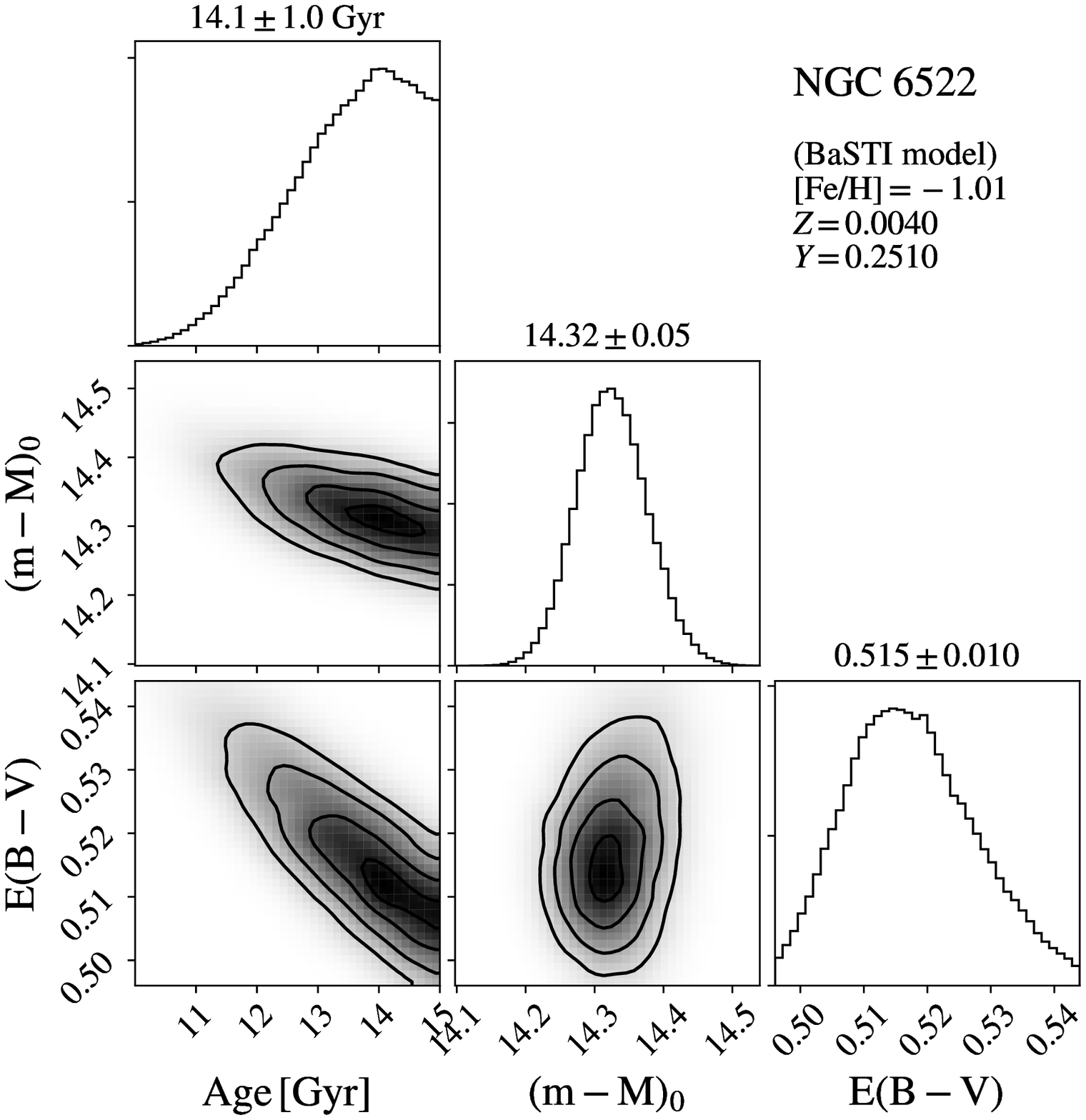}
\caption{
  Left panels: the observed CMD and the fiducial line for NGC\,6522 compared with a
  synthetic CMD for the best isochrone fit using $\alpha$-enhanced BaSTI models with [Fe/H]$=-1.01$ and
  $Y=0.251$.
  Isochrones with 11.0, 12.0, 13.0, 14.0, and 15.0 Gyr are overplotted
  (black lines). In each panel, cluster members (left) and synthetic stars (right) are shown  (gray points).
  The color difference between the data and the fiducial line for the best fit
  is also presented.   
  Right panels: corner plots showing the output from the MCMC method.
  They present the one- and two-dimensional projections
  of the posterior probability distributions for all parameters.
  The contours correspond to the [0.5$\sigma$, 1$\sigma$, 1.5$\sigma$, 2$\sigma$] levels.
}
\centering
\label{NGC6522_cmd_obs_vs_models_BaSTI_fehm115}
\end{figure*}

Figure \ref{NGC6522_cmd_obs_vs_models_BaSTI_fehm115} illustrates our
isochrone fitting method applied to the \textit{HST} CMD of NGC\,6522.
In this fit we employed $\alpha$-enhanced BaSTI models with [Fe/H]$=-1.01$ and $Y=0.251$.
The left panels in this figure present the observed CMD and its fiducial line,
as well as their synthetic counterparts for the best solution,
whereas the right panel shows the MCMC sampling results for NGC\,6522.
The complete set of results for all clusters, including all figures with the output of the MCMC,
is available in the online material.

\section{Results from isochrone fitting}
\label{results}

\subsection{NGC\,6522}

\begin{figure*}
\centering
\includegraphics[angle=0,scale=0.40]{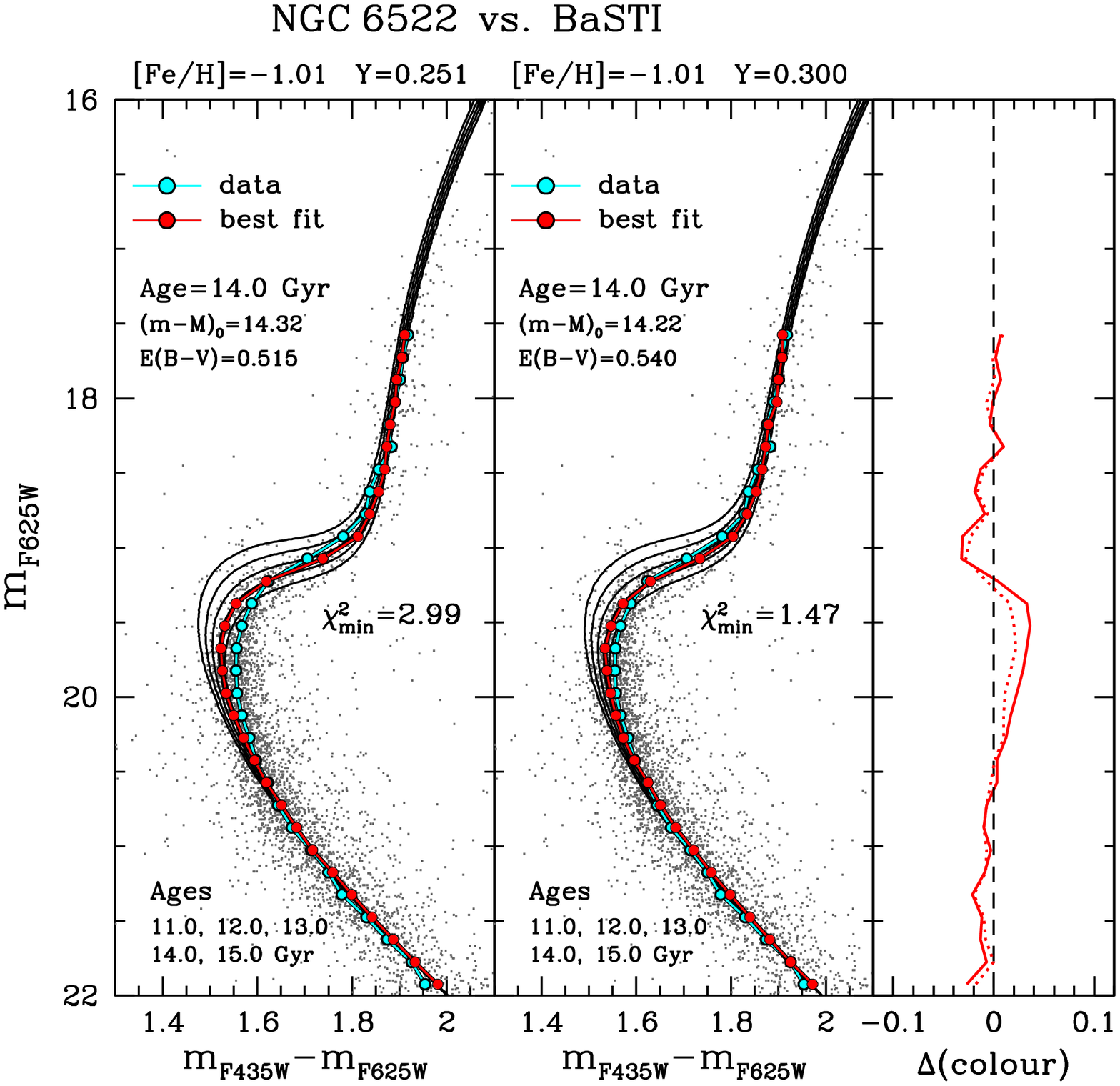}
\includegraphics[angle=0,scale=0.40]{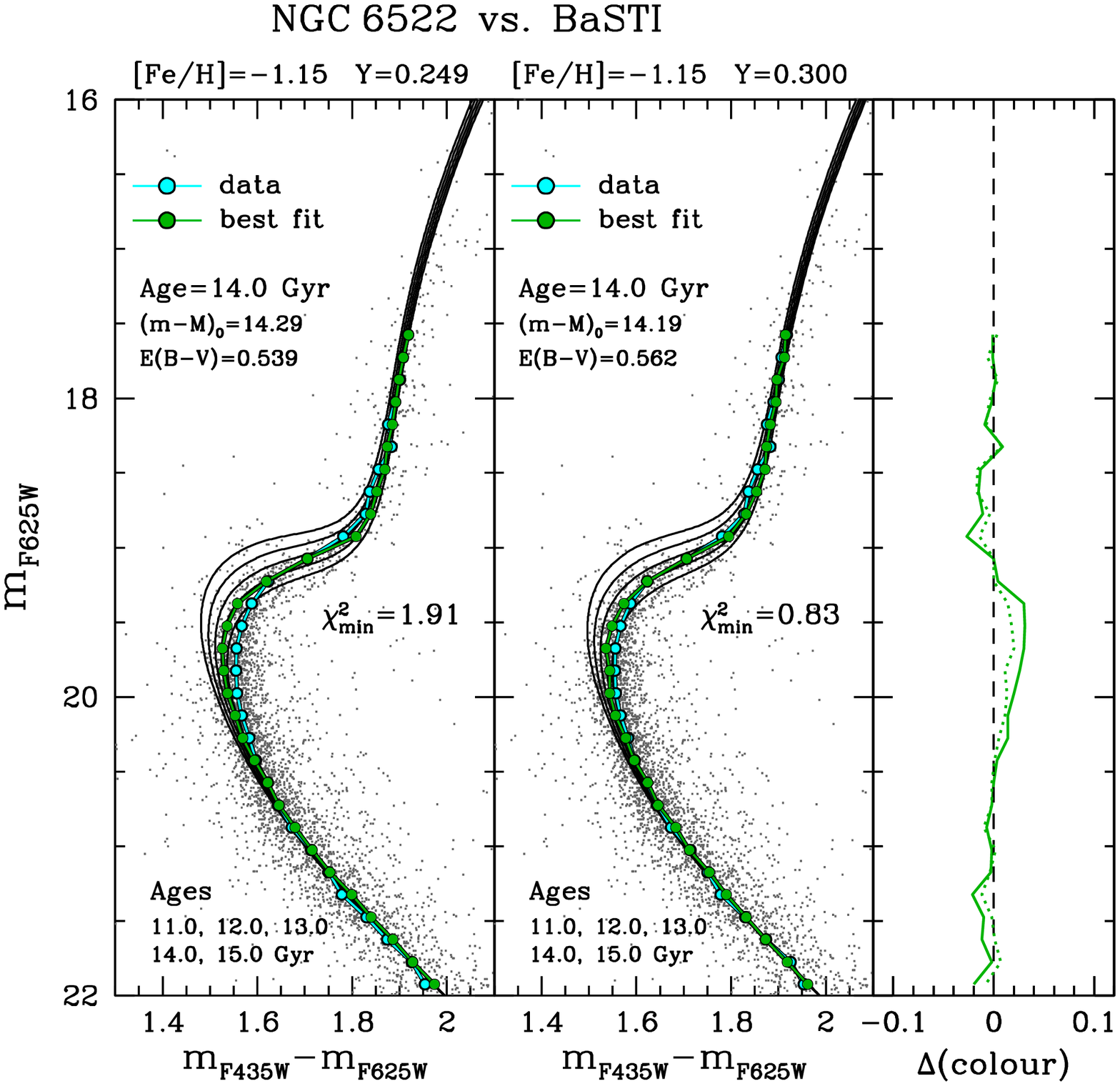}
\caption{
  Left panels: the observed fiducial line for NGC\,6522 compared with a
  synthetic one for the best isochrone fit using $\alpha$-enhanced BaSTI models with [Fe/H]$=-1.01$.
  Two different helium abundances were tested: $Y\sim 0.25$ and $Y=0.30$.
  Isochrones with 11.0, 12.0, 13.0, 14.0, and 15.0 Gyr are overplotted
  (black lines). The cluster members are also shown (gray points).
  The color differences between the data and the best fiducial lines for models with $Y\sim 0.25$ (solid line)
  and $Y=0.30$ (dotted line) are also presented.
  Right panels: the same as in the left panels, but for the BaSTI models with [Fe/H]$=-1.15$.}
\centering
\label{NGC6522_cmd_obs_vs_models_BaSTI}
\end{figure*}

\begin{figure*}
\centering
\includegraphics[angle=0,scale=0.40]{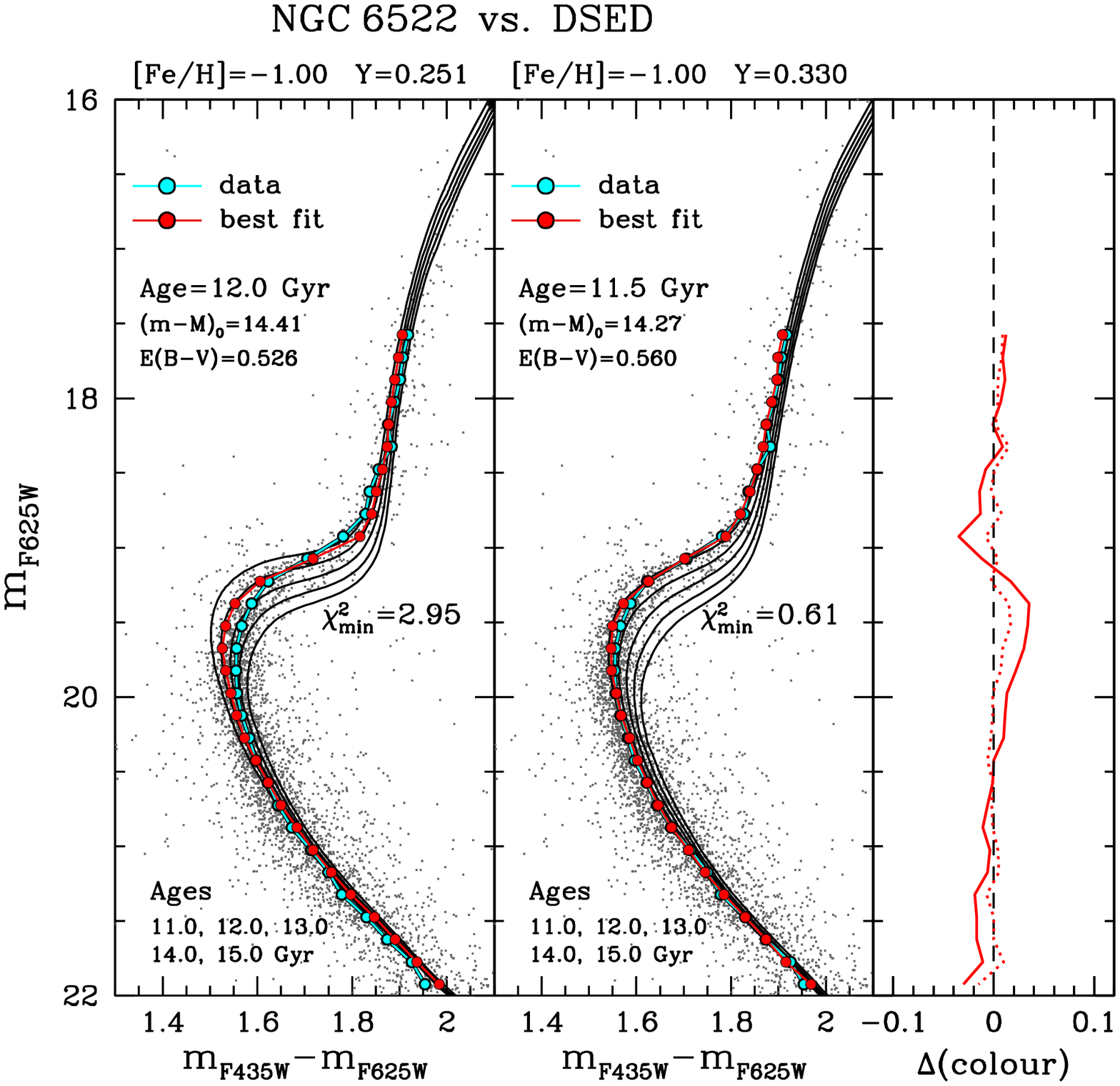}
\includegraphics[angle=0,scale=0.40]{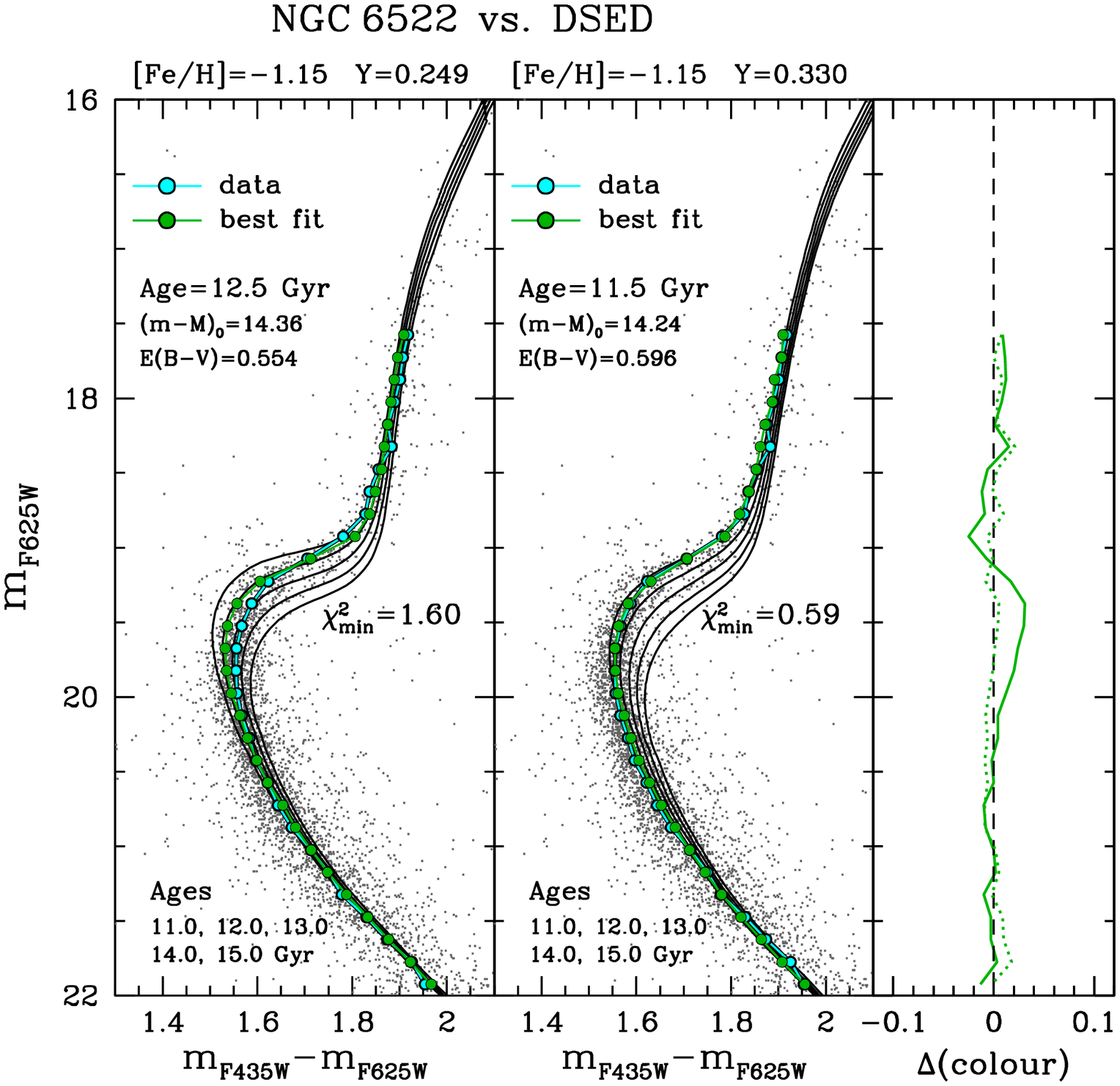}
\caption{Same as Figure \ref{NGC6522_cmd_obs_vs_models_BaSTI},
  but for DSED isochrones with [Fe/H]$=-1.00$ (left panels) and [Fe/H]$=-1.15$ (right panels).
  The DSED helium-enhanced models have $Y=0.33$.}
\centering
\label{NGC6522_cmd_obs_vs_models_DSED}
\end{figure*}

For this cluster two different values of metallicity were used:
[Fe/H]=$-$1.0 and $-$1.15.
The best isochrone fits using BaSTI and DSED models are shown in
Figs. \ref{NGC6522_cmd_obs_vs_models_BaSTI} and \ref{NGC6522_cmd_obs_vs_models_DSED}.
The recovered physical parameters are presented in these figures and in Table \ref{tab_all_parameters},
including the minimum $\chi^{2}$ for each set of models ($\chi^{2}_{\rm{min}}$). 

The results with BaSTI isochrones indicate an age of $\sim$ 14.2 Gyr for this cluster,
or $\sim$ 13.3 Gyr when corrected for the effect of atomic diffusion (See Section 5.1).
On the other hand, DSED models point to $\sim$ 12.2 Gyr if a canonical helium abundance ($Y\sim 0.25$)
is employed, or $\sim$ 11.4 Gyr if  a $\Delta Y \sim 0.08$ is considered.
The age results seem to be insensitive to the choice of metallicity and present
random uncertainties that are $\sim$ 1.0 Gyr. 
In contrast, there are significant variations in distance modulus and reddening as a function
of the adopted stellar evolutionary model and helium abundance.
As expected, helium-enhanced models recovered solutions with shorter distances and higher reddening values,
producing systematic differences of $\sim$ 0.13 mag in distance modulus ($\sim 0.4$ kpc) and 
$\sim$ 0.04 in $E$(\textit{B}$-$\textit{V}).
Besides this, in comparison to results from BaSTI models, DSED isochrones favored
isochrone fits with significantly higher values for distance and reddening.

Another remarkable result concerns the dependency of the goodness of fit with the helium abundance:
as attested by the color difference between model and data and consequently by the $\chi^2_{\rm{min}}$ value,
the shape of the observed fiducial line in the MSTO region and  the SGB region seems to be better
reproduced by helium-enhanced models, independently of the choice related to the stellar evolutionary model.
However, in Section 7.2 we demonstrate that the helium enhancement hypothesis is ruled out by the
mean magnitude of the RR Lyrae stars.

\begin{table*}
\begin{center}
  \caption{Physical Parameters for NGC\,6522, NGC\,6626, and NGC\,6362 from Isochrone Fits Using $\alpha$-enhanced BaSTI and DSED Models}
\begin{tabular}{lcccccccccc}
\tableline
\tableline
\hline
\tableline
Cluster & Model  & [Fe/H] & $Z$ & $Y$ & Age (Gyr)  & $(m-M)_{0}$  & $d_{\odot}$ (kpc) & $R_{\rm{GC}}$ & $E$(\textit{B}$-$\textit{V}) & $\chi^{2}_{\rm{min}}$ \\ 
\tableline
\tableline
NGC\,6522 & BaSTI & -1.01 & 0.0040 & 0.2510 & 14.1 $\pm$ 1.0$^{a}$ & 14.32 $\pm$ 0.05 & 7.31 $\pm$ 0.17 & 0.88 $\pm$ 0.12 & 0.515 $\pm$ 0.010 & 2.99 \\
 & BaSTI & -1.01 & 0.0037 & 0.3000 & 14.1 $\pm$ 1.0$^{a}$ & 14.22 $\pm$ 0.05 & 6.98 $\pm$ 0.16 & 1.15 $\pm$ 0.14 & 0.540 $\pm$ 0.009 & 1.47 \\
 & BaSTI & -1.15 & 0.0029 & 0.2490 & 14.2 $\pm$ 0.9$^{a}$ & 14.29 $\pm$ 0.05 & 7.21 $\pm$ 0.17 & 0.96 $\pm$ 0.13 & 0.539 $\pm$ 0.009 & 1.91 \\
 & BaSTI & -1.15 & 0.0027 & 0.3000 & 14.4 $\pm$ 1.0$^{a}$ & 14.19 $\pm$ 0.05 & 6.89 $\pm$ 0.16 & 1.23 $\pm$ 0.14 & 0.562 $\pm$ 0.010 & 0.83 \\
 \tableline
 & DSED & -1.00 & 0.0035 & 0.2506 & 12.1 $\pm$ 1.0 & 14.41 $\pm$ 0.05 & 7.62 $\pm$ 0.18 & 0.67 $\pm$ 0.08 & 0.526 $\pm$ 0.013 & 2.95 \\
 & DSED & -1.00 & 0.0031 & 0.3300 & 11.4 $\pm$ 1.0 & 14.27 $\pm$ 0.05 & 7.14 $\pm$ 0.17 & 1.01 $\pm$ 0.14 & 0.560 $\pm$ 0.011 & 0.61 \\
 & DSED & -1.15 & 0.0023 & 0.2488 & 12.4 $\pm$ 1.0 & 14.36 $\pm$ 0.05 & 7.45 $\pm$ 0.17 & 0.78 $\pm$ 0.10 & 0.554 $\pm$ 0.011 & 1.60 \\
 & DSED & -1.15 & 0.0021 & 0.3300 & 11.4 $\pm$ 1.0 & 14.24 $\pm$ 0.05 & 7.05 $\pm$ 0.16 & 1.09 $\pm$ 0.13 & 0.596 $\pm$ 0.013 & 0.59 \\
\tableline
\tableline 
NGC\,6626 & BaSTI & -1.31 & 0.0020 & 0.2480 & 14.3 $\pm$ 1.0$^{a}$ & 13.60 $\pm$ 0.06 & 5.25 $\pm$ 0.15 & 2.96 $\pm$ 0.13 & 0.424 $\pm$ 0.010 & 1.61 \\
(M28) & BaSTI & -1.31 & 0.0019 & 0.3000 & 14.0 $\pm$ 1.1$^{a}$ & 13.51 $\pm$ 0.06 & 5.04 $\pm$ 0.14 & 3.15 $\pm$ 0.13 & 0.447 $\pm$ 0.010 & 0.58 \\
\tableline
 & DSED & -1.30 & 0.0016 & 0.2477 & 12.1 $\pm$ 1.0 & 13.67 $\pm$ 0.06 & 5.42 $\pm$ 0.15 & 2.81 $\pm$ 0.13 & 0.440 $\pm$ 0.013 & 1.28 \\
 & DSED & -1.30 & 0.0015 & 0.3300 & 11.1 $\pm$ 0.9 & 13.57 $\pm$ 0.06 & 5.18 $\pm$ 0.14 & 3.02 $\pm$ 0.12 & 0.476 $\pm$ 0.011 & 0.39 \\
\tableline 
\tableline
NGC\,6362 & BaSTI & -1.01 & 0.0040 & 0.2510 & 14.3 $\pm$ 1.0$^{a}$ & 14.40 $\pm$ 0.05 & 7.59 $\pm$ 0.17 & 5.11 $\pm$ 0.05 & 0.038 $\pm$ 0.011 & 1.54 \\
 & BaSTI & -1.01 & 0.0037 & 0.3000 & 14.3 $\pm$ 1.1$^{a}$ & 14.30 $\pm$ 0.05 & 7.24 $\pm$ 0.17 & 5.04 $\pm$ 0.03 & 0.060 $\pm$ 0.012 & 1.17 \\
 & BaSTI & -1.15 & 0.0029 & 0.2490 & 14.7 $\pm$ 0.9$^{a}$ & 14.38 $\pm$ 0.05 & 7.52 $\pm$ 0.17 & 5.02 $\pm$ 0.03 & 0.062 $\pm$ 0.011 & 1.09 \\
 & BaSTI & -1.15 & 0.0027 & 0.3000 & 14.8 $\pm$ 1.1$^{a}$ & 14.28 $\pm$ 0.05 & 7.18 $\pm$ 0.17 & 5.02 $\pm$ 0.03 & 0.091 $\pm$ 0.011 & 1.58 \\
\tableline
 & DSED & -1.00 & 0.0035 & 0.2506 & 12.8 $\pm$ 1.0 & 14.48 $\pm$ 0.05 & 7.87 $\pm$ 0.18 & 5.19 $\pm$ 0.06 & 0.040 $\pm$ 0.015 & 0.99 \\
 & DSED & -1.00 & 0.0031 & 0.3300 & 11.2 $\pm$ 1.2 & 14.36 $\pm$ 0.05 & 7.45 $\pm$ 0.17 & 5.08 $\pm$ 0.04 & 0.081 $\pm$ 0.016 & 2.09 \\
 & DSED & -1.15 & 0.0023 & 0.2488 & 12.8 $\pm$ 1.0 & 14.44 $\pm$ 0.05 & 7.73 $\pm$ 0.18 & 5.15 $\pm$ 0.05 & 0.070 $\pm$ 0.014 & 0.96\\
 & DSED & -1.15 & 0.0021 & 0.3300 & 11.2 $\pm$ 1.1 & 14.32 $\pm$ 0.05 & 7.31 $\pm$ 0.17 & 5.05 $\pm$ 0.04 & 0.128 $\pm$ 0.016 & 2.56\\
\tableline 

\tableline 
\end{tabular}
\end{center}
\tablecomments{
 $^{a}$Ages using BaSTI models neglecting the effect of atomic diffusion.
  The inclusion of this effect reduces the ages by about 0.9 Gyr
  (See Section \ref{isochronemodels} for more details).
}
\label{tab_all_parameters}
\end{table*}

\subsection{NGC\,6626}

\begin{figure*}
\centering
\includegraphics[angle=0,scale=0.40]{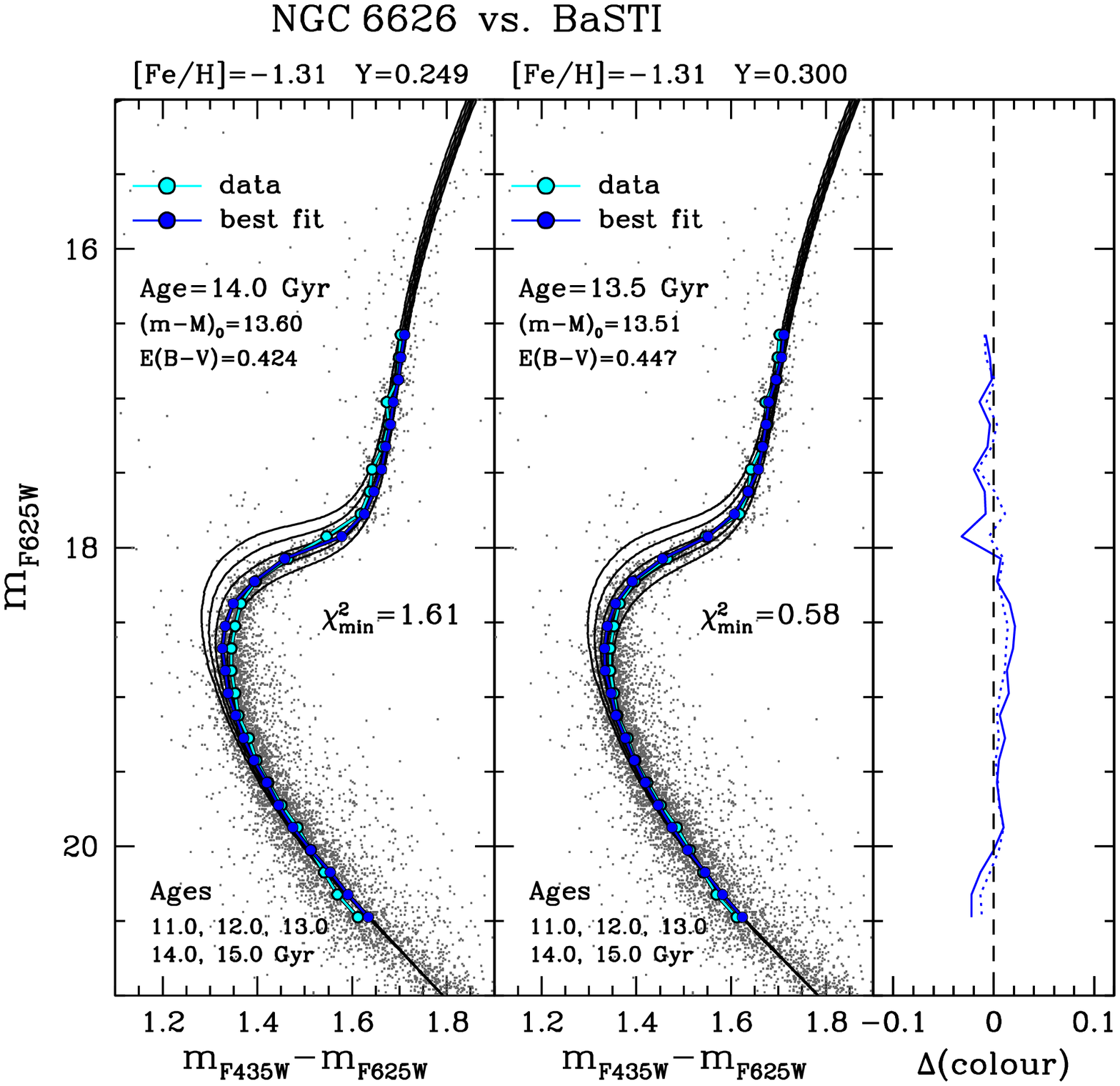}
\includegraphics[angle=0,scale=0.40]{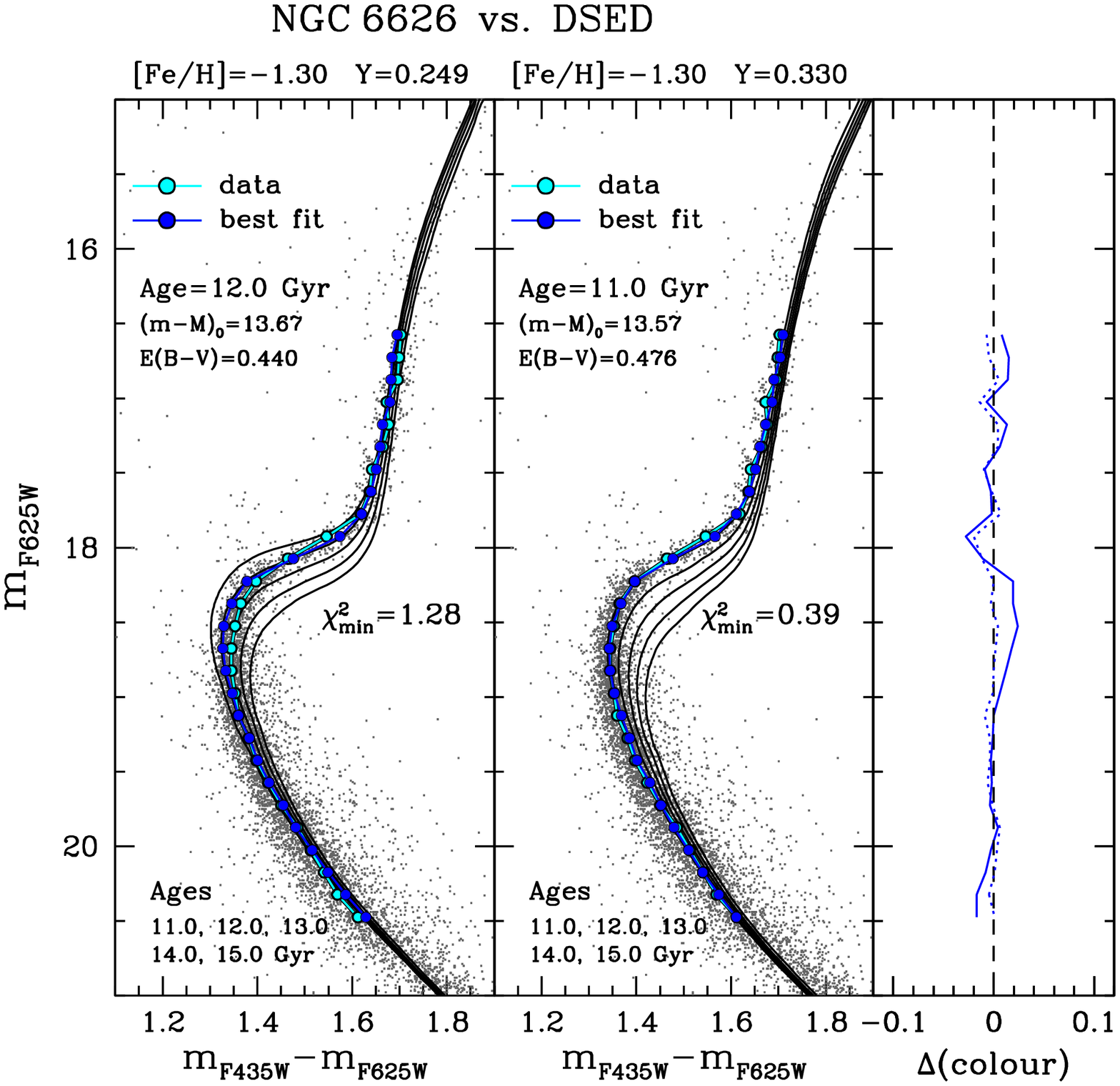}
\caption{Same as Figure \ref{NGC6522_cmd_obs_vs_models_BaSTI},
  but for NGC\,6626 and BaSTI (left panels) and DSED (right panels) isochrones with [Fe/H]$\sim-1.3$.}
\centering
\label{NGC6626_cmd_obs_vs_models_fehm130}
\end{figure*}

Figure \ref{NGC6626_cmd_obs_vs_models_fehm130}
presents the observed fiducial line for NGC\,6626 in comparison with the best isochrone fits
using DSED and BaSTI stellar evolutionary models.
We adopted a single metallicity for this cluster, very close to the
 value determined by \citet{Villanova+17}.
As demonstrated by the previous results related to NGC\,6522, this constraint is crucial to reduce the systematic uncertainties
in distance and reddening. The recovered physical parameters are shown in Table \ref{tab_all_parameters}.

This cluster presents very similar results with respect to NGC\,6522.
Assuming the same helium abundance and the same stellar evolutionary model,
their ages are almost identical.
As occurred for NGC\,6522, the goodness of fit increases significantly with the adoption of helium-enhanced models.
However, as demonstrated in Section 7.2, once again a helium enhancement of $\Delta Y\sim 0.05$ is not
compatible with the mean magnitude of RR Lyrae stars observed in this cluster.

\subsection{NGC\,6362}

\begin{figure*}
\centering
\includegraphics[angle=0,scale=0.40]{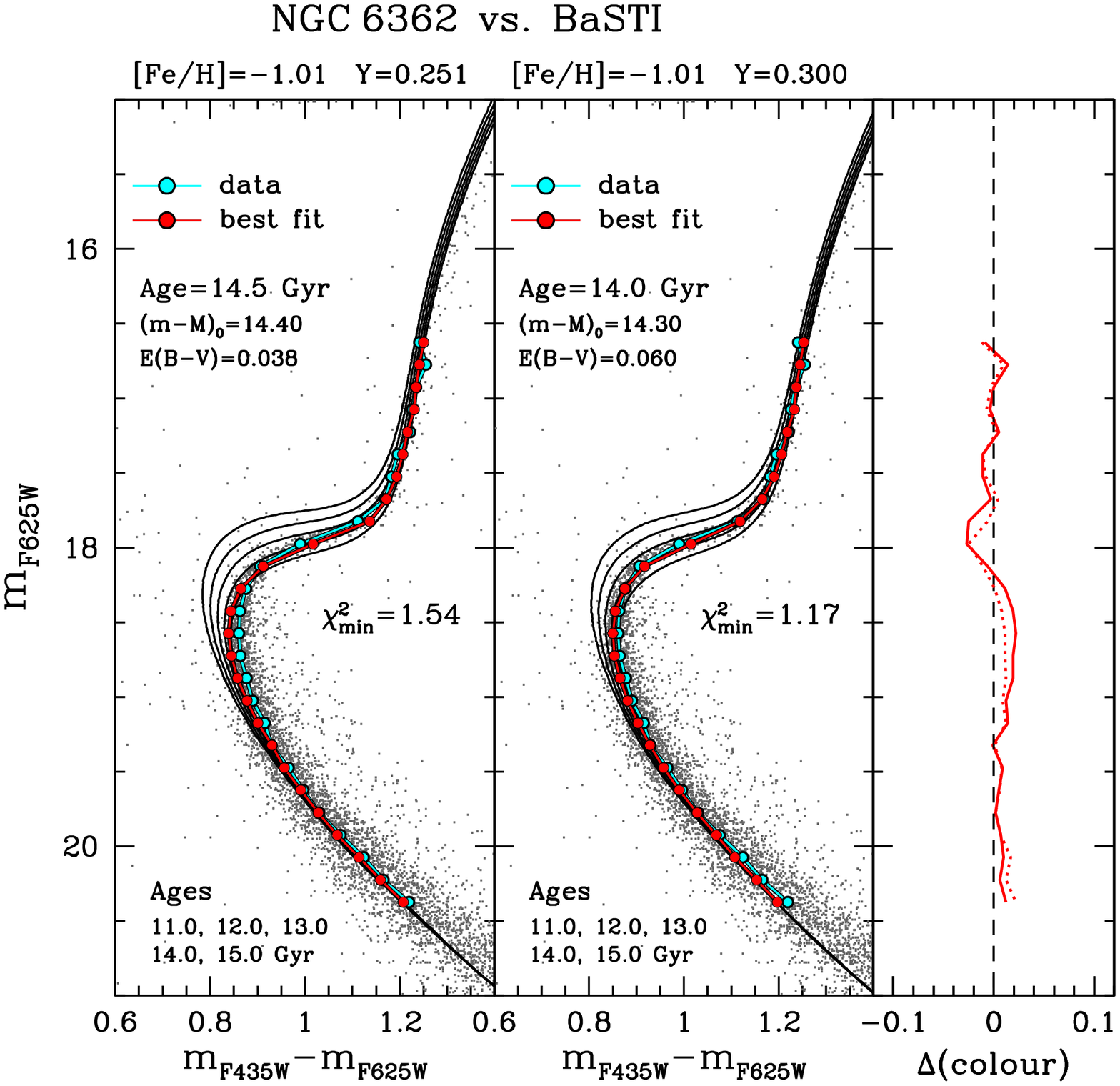}
\includegraphics[angle=0,scale=0.40]{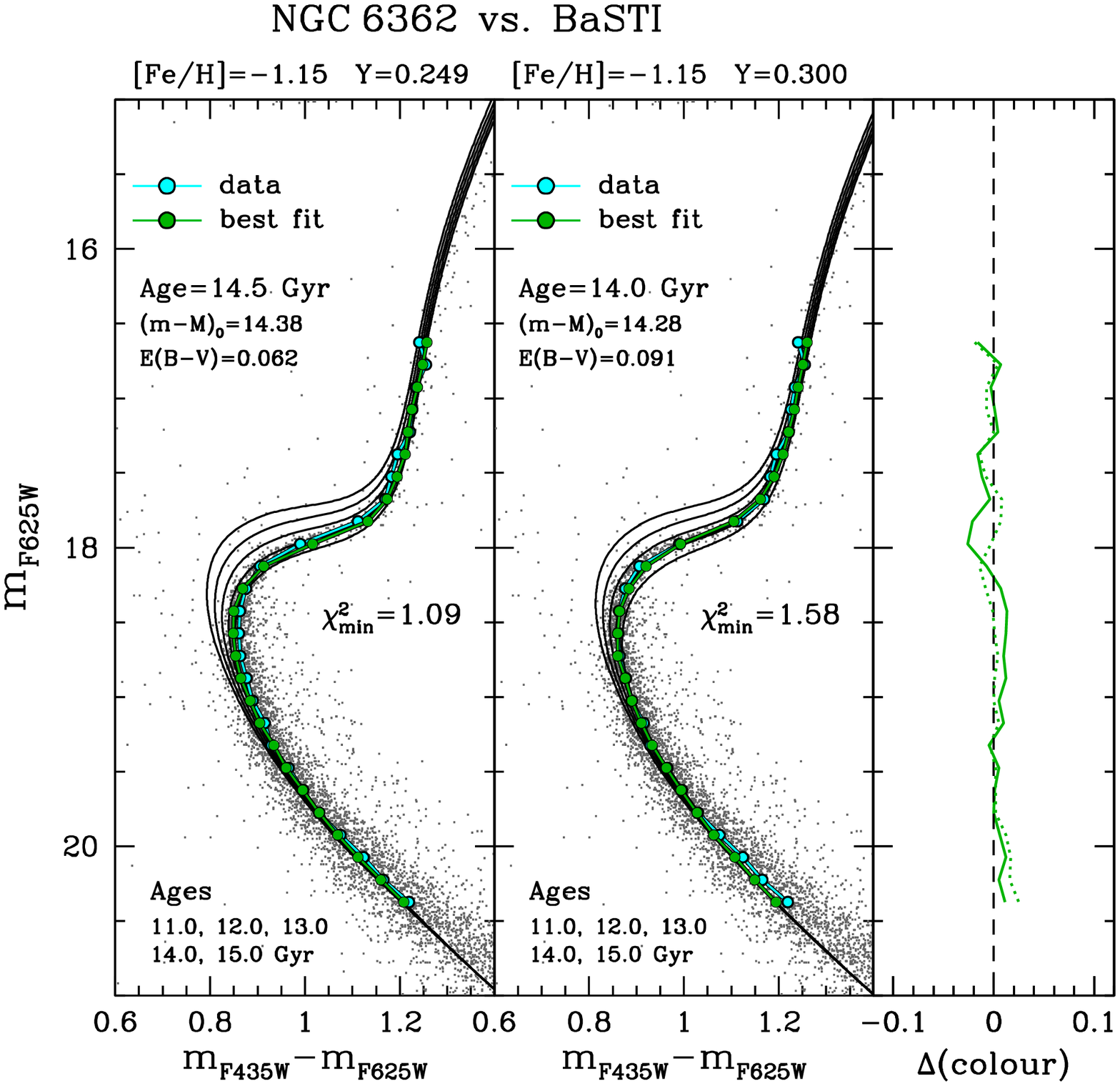}
\caption{Same as  Figure \ref{NGC6522_cmd_obs_vs_models_BaSTI}, 
  but for NGC\,6362.}
\centering
\label{NGC6362_cmd_obs_vs_models_BaSTI}
\end{figure*}

\begin{figure*}
\centering
\includegraphics[angle=0,scale=0.40]{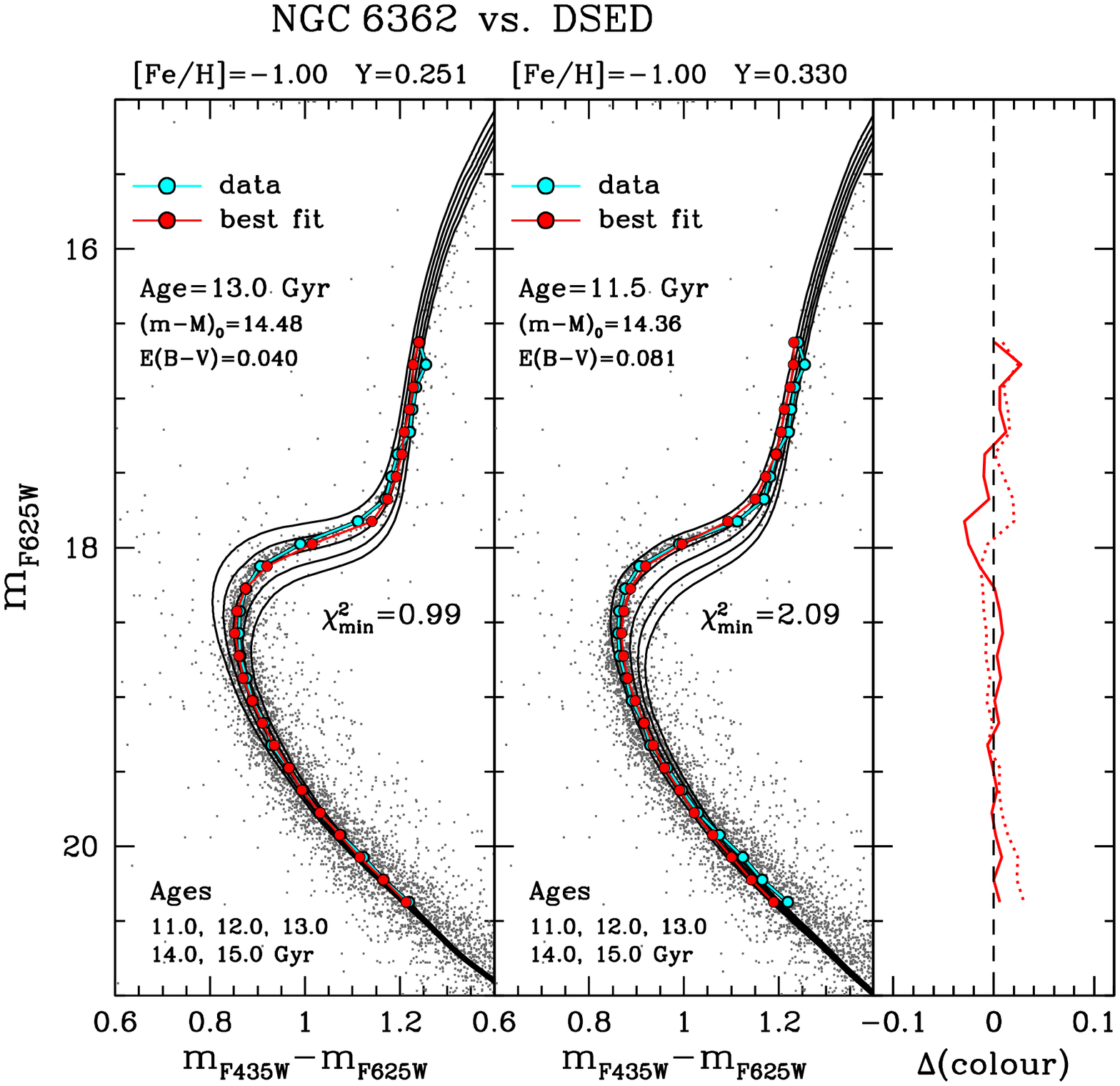}
\includegraphics[angle=0,scale=0.40]{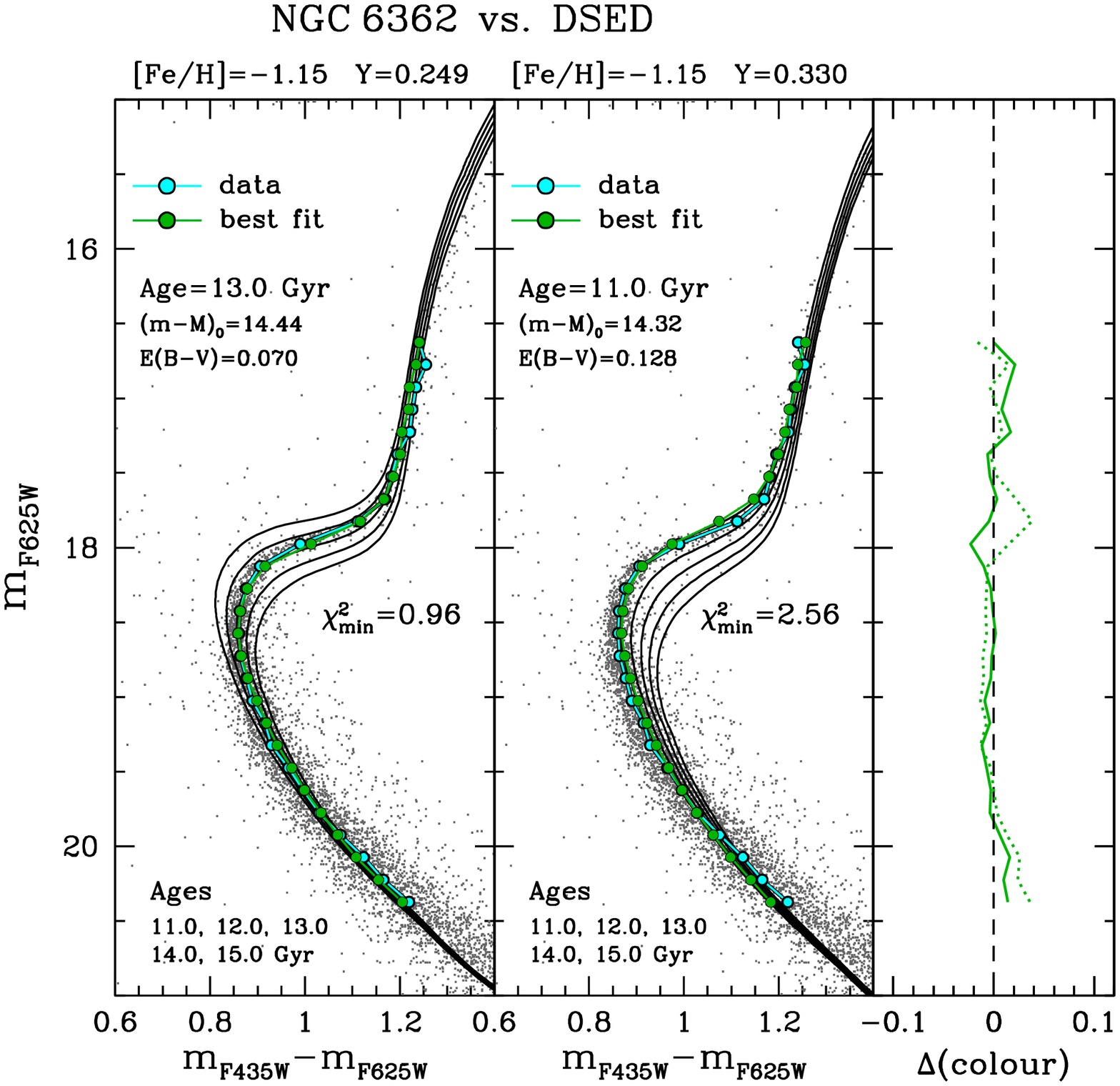}
\caption{Same as  Figure \ref{NGC6522_cmd_obs_vs_models_DSED},
  but for NGC\,6362.}
\centering
\label{NGC6362_cmd_obs_vs_models_DSED}
\end{figure*}

The eight isochrone fits for this cluster using different combinations of metallicity,
helium abundance, and stellar evolutionary models are presented in Figs. 
\ref{NGC6362_cmd_obs_vs_models_BaSTI} and
\ref{NGC6362_cmd_obs_vs_models_DSED}.
The physical parameters determined in this analysis are shown in Table \ref{tab_all_parameters}.

Regardless of the choice in metallicity and in helium abundance,
the BaSTI models also indicated an old age for this cluster
($\sim$ 13.5 Gyr/14.5 Gyr with/without atomic diffusion).
This evidence suggests a slightly older age in comparison with the other two clusters, but
still in agreement if the uncertainties ($\sim$ 1.0 Gyr) are taken into account.
The sample clusters are found to be coeval also from the results for DSED models,
within uncertainties, with ages of $\sim$ 12.4 Gyr assuming a canonical helium abundance
($Y\sim 0.25$) or $\sim$ 11.2 Gyr for $Y\sim 0.33$.

In contrast with the results for NGC\,6522 and NGC\,6626, 
Table \ref{tab_all_parameters} shows that the goodness of fit for NGC\,6362 does
not favor helium enhancement using BaSTI models, and for DSED models, a canonical helium
is clearly favored. 
Assuming a helium abundance of $Y\sim 0.25$, and a metallicity taken as the average between
the two most recent determinations ([Fe/H]$=-1.08$; 
\citealt{Mucciarelli+16, Massari+17}), we get an intrinsic
distance modulus of 14.39 (BaSTI) or 14.46 (DSED)
and  $E$(\textit{B}$-$\textit{V}) of 0.050 (BaSTI) or 0.055 (DSED).

\section{Discussion}

\begin{figure*}
\centering
\includegraphics[angle=0,scale=.65]{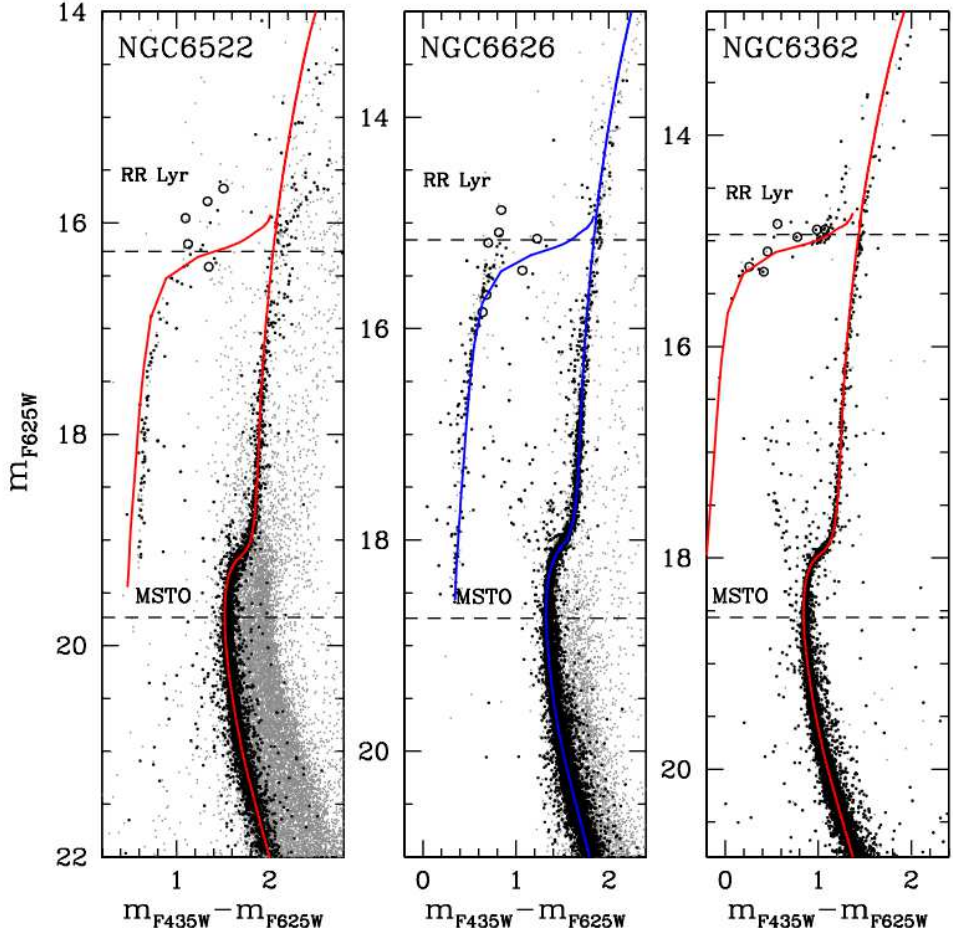}
\caption{F626W vs. F435W--F625W CMDs for the three sample clusters.
BaSTI isochrones of 14 Gyr, $Y=0.25$ are shown as red sequences corresponding to
[Fe/H]$=-1.01$ and blue ones corresponding to [Fe/H]$=-1.31$. 
The blue ZAHB sequence corresponds to tracks with
($Z=0.003$, $Y=0.251$) and the red ZAHB sequence corresponds to 
 ($Z=0.004$, $Y=0.251$). Dashed lines represent the MSTO and HB levels (as in Figure \ref{cmds}).
}
\label{CMDfinal} 
\end{figure*}

\begin{figure*}
\centering
\includegraphics[angle=0,scale=.29]{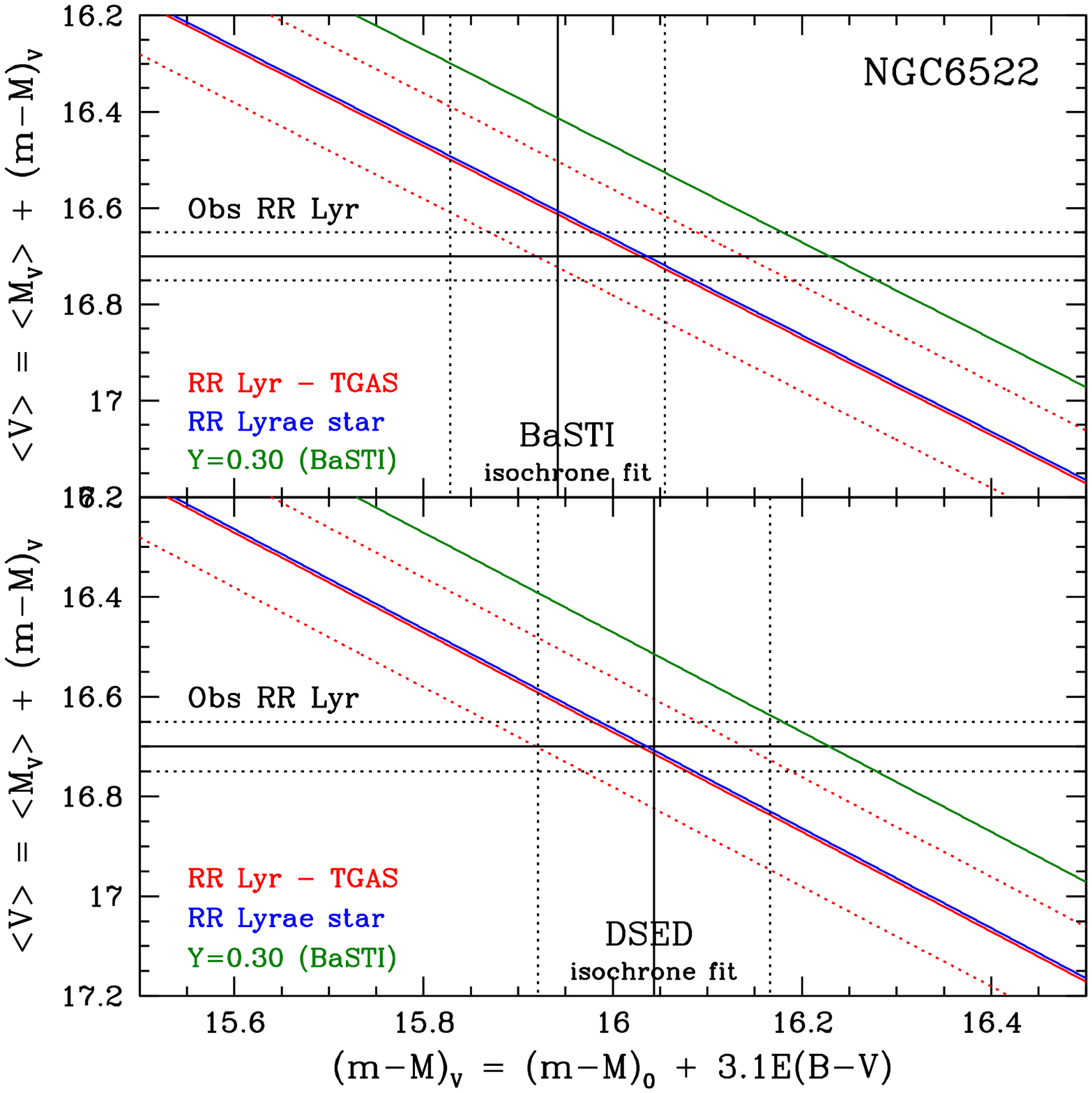}
\includegraphics[angle=0,scale=.29]{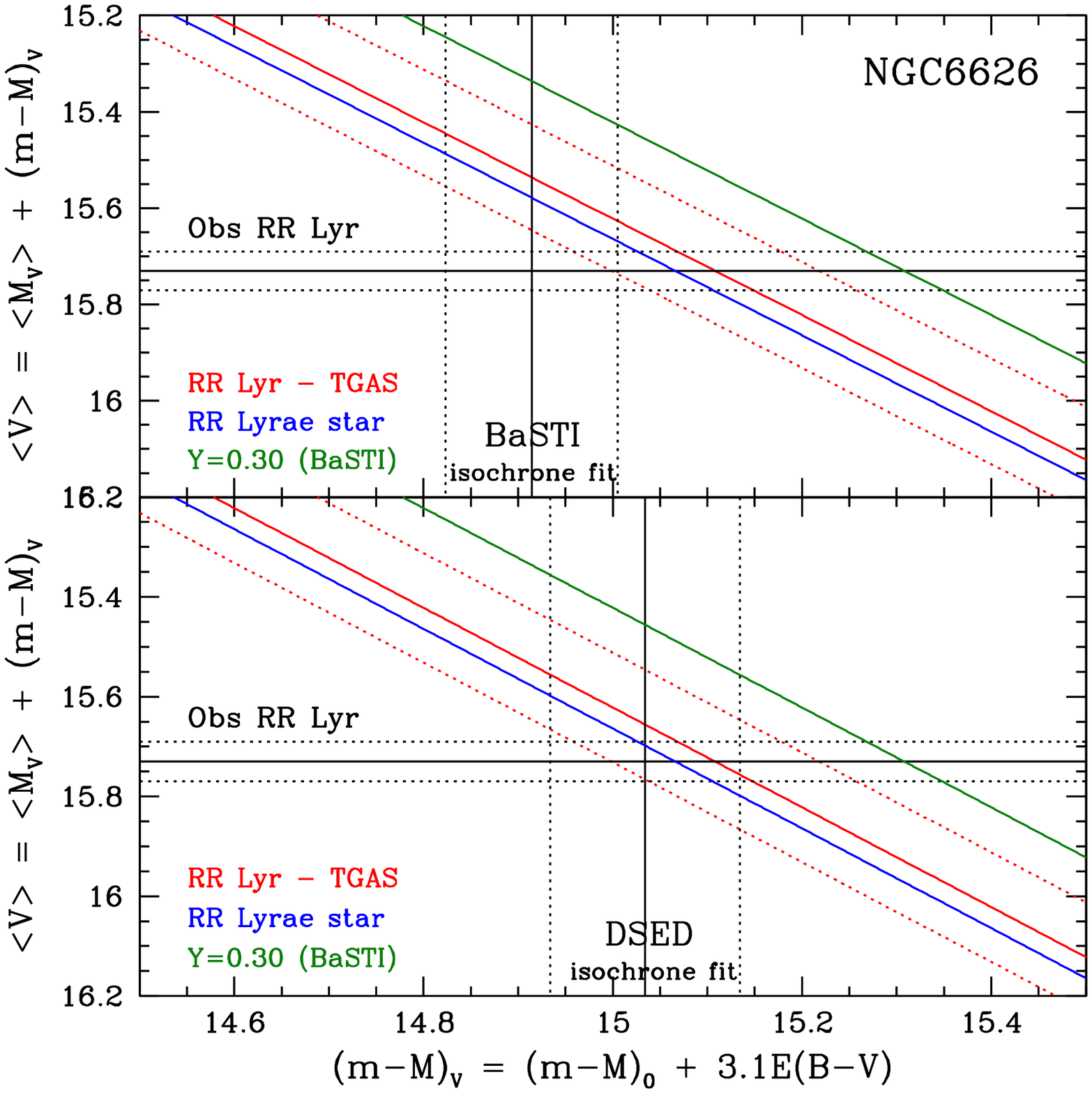}
\includegraphics[angle=0,scale=.29]{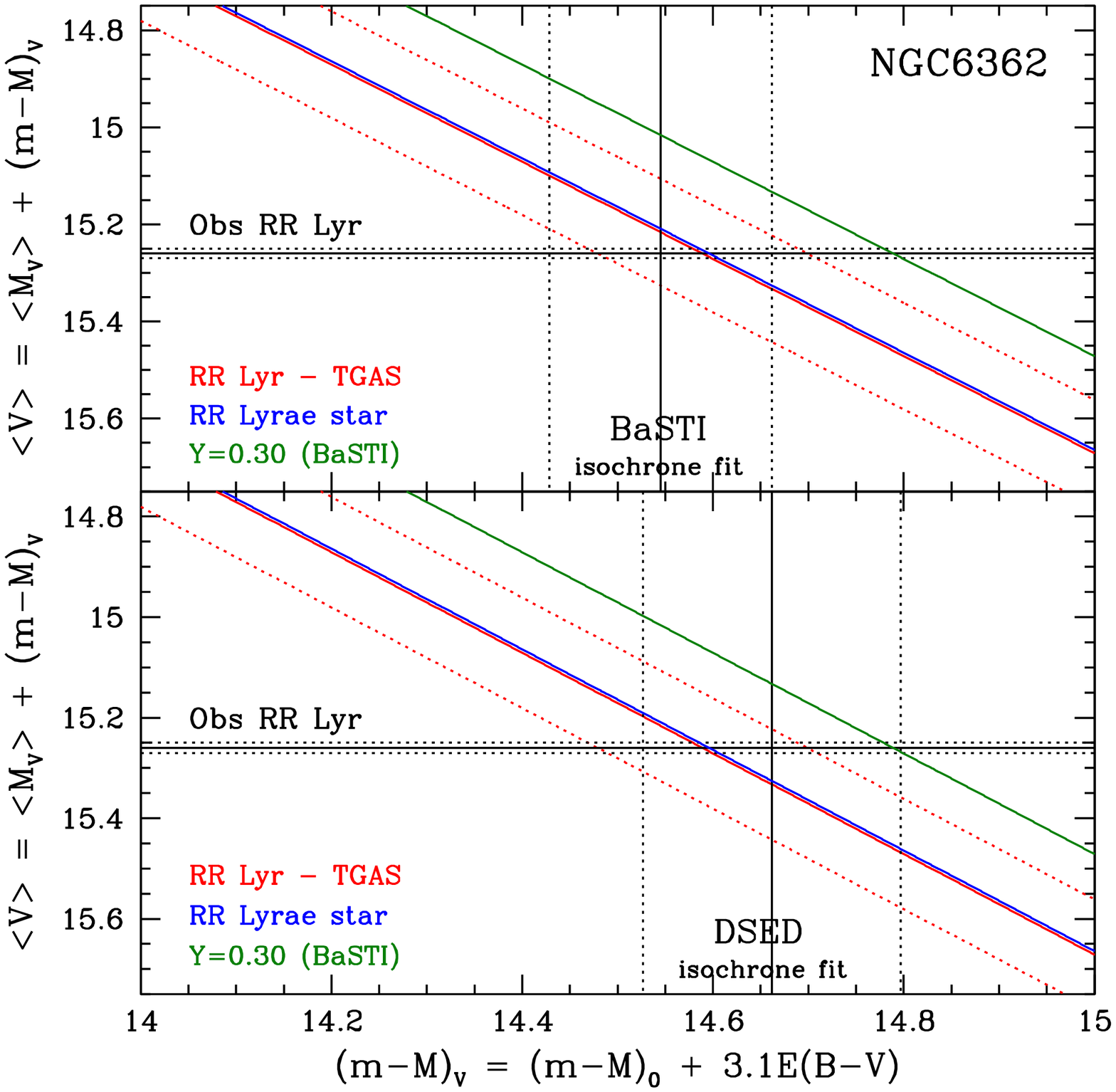}
\caption{Predicted average $V$ magnitude for the RR Lyrae stars as a function of the apparent distance modulus in $V$
  for NGC\,6522 (left panels), NGC\,6626 (middle panels), and NGC\,6362 (right panels).
  The analysis is done over two empirical $\langle M_{\rm{v}} \rangle$ RR Lyrae values:
  the recent one obtained by the \cite{Gaia17}
  (red solid lines, with 1$\sigma$ confidence level marked by red dotted lines),
  and the one for the RR Lyrae star itself (blue solid line; \citealp{Catelan+Cortes08}).  
  The results for the apparent distance modulus in $V$ from the isochrone fits using BaSTI (top panels) and DSED (bottom panels) are
  presented (vertical solid lines) within 1$\sigma$ confidence level (dotted lines).
  The average of the mean RR Lyrae $V$ magnitudes and its standard deviation are depicted in each panel
  (horizontal solid and dotted lines, respectively).
  The expected increment of $-0.20$ mag in $V$ predicted by the BaSTI models due to a helium enhancement of $\Delta Y\sim0.05$
  in the canonical helium abundance for a cluster with [Fe/H]$\sim-1.15$ ($Y\sim 0.25$) is also shown (green solid lines).
}
\label{Vmag_RRLyrae} 
\end{figure*}

Figure \ref{CMDfinal} summarizes the best isochrone fits  to the 
MSTO, SGB, and RGB from the statistical analysis, together with the newly
calculated tracks for the ZAHB, with all models from BaSTI.
For NGC\,6626 and NGC\,6362 the fits are remarkably good all along the sequences.
For NGC\,6522 the SGB-RGB sequences are well fitted, although there is
a slight offset around the MSTO. The theoretical ZAHB location is 
bluer/brighter than the observed distribution. We postpone a detailed 
investigation of this issue to a forthcoming
paper; however, we note that a variation (increase) of the adopted 
metallicity (see data in Table \ref{tab_spec}) would help in improving the quality of 
the fit. 
If a higher helium is adopted (e.g., $Y=0.30$), the HB models have
a higher luminosity by $\sim$0.25 mag, at the color of the RR Lyrae.

\subsection{Comparisons with literature}

From a comparison between the present results
and literature results  (Table \ref{tab_param_lit}), 
we confirm an age $\gtrsim 12.0$ Gyr for all clusters under the assumption 
of a canonical helium abundance.
A very old age of $\sim$ 14.0 Gyr
(or 13 Gyr if corrected for diffusion)
 for NGC\,6522 and NGC\,6626 obtained by the preceding studies
using \textit{HST} data \citep{Testa+01, Meissner+Weiss06, Barbuy+09}
is in good agreement with our results using BaSTI models.
Note that \citet{Testa+01} found evidence for a canonical
helium abundance in NGC\,6626.

The ages obtained for NGC\,6362 using BaSTI models ($\sim$ 14.0 Gyr) are compatible with the ones from 
\citet{Marin-Franch+09} and \citet{Paust+10}, 
whereas the ages $\sim$ 13.0 Gyr from our fits using DSED models,
and BaSTI models if diffusion is taken into account, in both cases
with a canonical helium abundance, are in better agreement with 
results from \citet{Dotter+10},
\cite{VandenBerg+13}, and \cite{Wagner-Kaiser+16}.
The distances and reddening values presented in our paper and those in the literature
 are comparable, but our results favor slightly shorter solar and
galactocentric distances and slightly lower $E$(\textit{B}$-$\textit{V}) values.

Ages of $\sim$12.5 Gyr are indicated by our fits with DSED isochrones for the three clusters
and derivations for other clusters using the same set of isochrones in the literature.
Our results also pointed out that these clusters are located at shorter solar distances
and have lower reddening values than previously thought.

Concerning NGC\,6522 structural parameters, we found a
 core radius of 12.32 $\pm$ 0.26 arcsec,
approximately four times ($\sim$ 4$\times$) higher than the one from
\cite{Trager+95} (3.1 arcsec), who classified this cluster as core collapse.
Due to the high stellar contamination and high reddening toward this cluster,
the analysis performed by \cite{Trager+95} over surface brightness profiles obtained from ground-based photometry 
likely underestimated the core radius for NGC\,6522.
Assuming the average solar distances recovered by us for NGC\,6522 (7.2 kpc), NGC\,6626 (5.2 kpc), and NGC\,6362 (7.5 kpc),
their intrinsic core radii are 0.43, 0.32, and 3.3 pc, respectively.
This reveals that NGC\,6626 is in fact the most compact cluster in our sample. 

\subsection{Constraints from RR Lyrae stars}

The RR Lyrae stars provide independent constraints on the apparent distance modulus, helium abundance, and age.
First, to bypass the theoretical limitations from the stellar evolutionary models and to avoid
using a conversion from $V$ to F625W, we analyzed the $V$ magnitude of the RR Lyrae stars directly from
Clement's catalog and the OGLE catalog, comparing them with two empirical average $\langle M_{V} \rangle$ determinations
for the RR Lyrae.
For this purpose we used the very recent $M_{V}-$[Fe/H] calibrated relations for RR Lyrae stars
from the \cite{Gaia17} based on Tycho-\textit{Gaia} Astrometric Solution (TGAS).
Since they employed three different methods to perform such calibration,
we assumed the mean $M_{V}$ solution for these approaches, taking into account three distinct sources of uncertainties:
(1) $\sim$ 0.10 dex in [Fe/H] (0.02 mag);
(2) the stochastic effects in the fits (0.04 mag);
(3) and the main one, the systematics due to the different solutions (0.10).
Assuming [Fe/H]$=-1.07 \pm 0.10$ as the metallicity of NGC\,6522 and NGC\,6362,
we obtained $\langle M_{V} \rangle = 0.67 \pm 0.11$ for their RR Lyrae stars.
Since NGC\,6626 is slightly more metal poor ([Fe/H]$=-1.30 \pm 0.10$),
such variable stars should be slightly brighter ($\langle M_{V} \rangle = 0.62 \pm 0.11$) in this cluster.
Due to the similarity with the metallicity of the RR Lyrae star itself ([Fe/H]$=-1.16$),
these values are very close to the one found for the prototype of this class of variable stars 
($\langle M_{V} \rangle = 0.66 \pm 0.14$, \citealp{Catelan+Cortes08}).

Figure \ref{Vmag_RRLyrae} presents the expected apparent $V$ magnitude for the RR Lyrae stars
as a function of the apparent distance modulus in $V$ (($m-M$)$_{0}$ + $A_V$) for all clusters analyzed in this work.
It is clear from these plots that the apparent distance moduli from the isochrone fits agree within 1$\sigma$
with what would be expected to bring the absolute magnitudes for the RR Lyrae stars to the observed ones. 
The only possible exception is the solution for NGC\,6626 using BaSTI models,
where a higher apparent distance modulus seems to be required.

We also checked whether the isochrone fits and the observed RR Lyrae stars can be consistent with helium-enhanced stars.
BaSTI models predict that RR Lyrae stars with $Y=0.30$ should be $\sim$ 0.20 mag brighter in $V$ than ones with a canonical
helium abundance for a metallicity of [Fe/H]$\sim -1.15$ ($Y\sim 0.25$).
The results for this simple experiment are also depicted in Figure \ref{Vmag_RRLyrae}, clearly revealing that such bright magnitudes
are not compatible with the mean $V$ magnitude of the RR Lyrae stars observed in these clusters at the expected
apparent distance moduli, therefore rejecting a helium enhancement of $\Delta Y\sim 0.05$.
In fact, the recent analysis of the red giant branch bump performed by \cite{Lagioia+18} on the data from the
\textit{HST} UV Legacy Survey of Galactic GCs support the hypothesis that the variations in the average helium abundance
between distinct subpopulations are lower than $\Delta Y\lesssim 0.03$.

Taking the advantage that all clusters have similar metallicities and their MSTO and RR Lyrae mean magnitudes were
homogeneously derived in the same filter, we computed the $\Delta V^{\rm{HB}}_{\rm{TO}} $  parameter in F625W to check whether there is
any indication of age differences between these clusters.
The results presented in Table \ref{tab_photom} reveal similar $\Delta V^{\rm{HB}}_{\rm{TO}} $  values within the uncertainties;
therefore, these clusters are probably coeval within $\sim$ 1.0 Gyr.
Further investigations to discover new RR Lyrae stars in the cores of NGC\,6522 and NGC\,6626,
as well as accurate MSTO determinations in the $V$ filter, can reduce the uncertainties in
$\Delta V^{\rm{HB}}_{\rm{TO}} $ and reveal some age difference.

\subsection{Testing the presence of multiple stellar populations}
\label{multiple}

The high quality of \textit{HST} data and proper-motion-cleaned CMDs
 allows us to investigate the
presence of multiple stellar populations (MPs) in NGC\,6522. 
This is the first time that this kind of analysis is employed in this cluster.

The presence of more than one stellar population in NGC\,6362
 was also detected by
combining optical \textit{HST} and UV ground-based photometry. \cite{Dalessandro+14}
have demonstrated that the RGB of NGC\,6362 is split into two separate sequences.
A deep photometric analysis performed by \cite{Piotto+15}
and \cite{Milone+17} using UV--optical \textit{HST} data,
clearly revealed the presence of these two stellar populations even
 for the MS stars. 
These results were recently corroborated by the high-resolution
spectroscopic analysis performed by \cite{Mucciarelli+16} and \cite{Massari+17}.

Concerning NGC\,6626, the signature of MPs was revealed by the abundance analysis of 17 RGB stars
from \cite{Villanova+17}, which presented an Na-O anticorrelation and an Na-Al correlation.
No photometric evidence of MPs in this cluster was found so far. 

\begin{figure*}
\centering
\includegraphics[angle=0,scale=0.42]{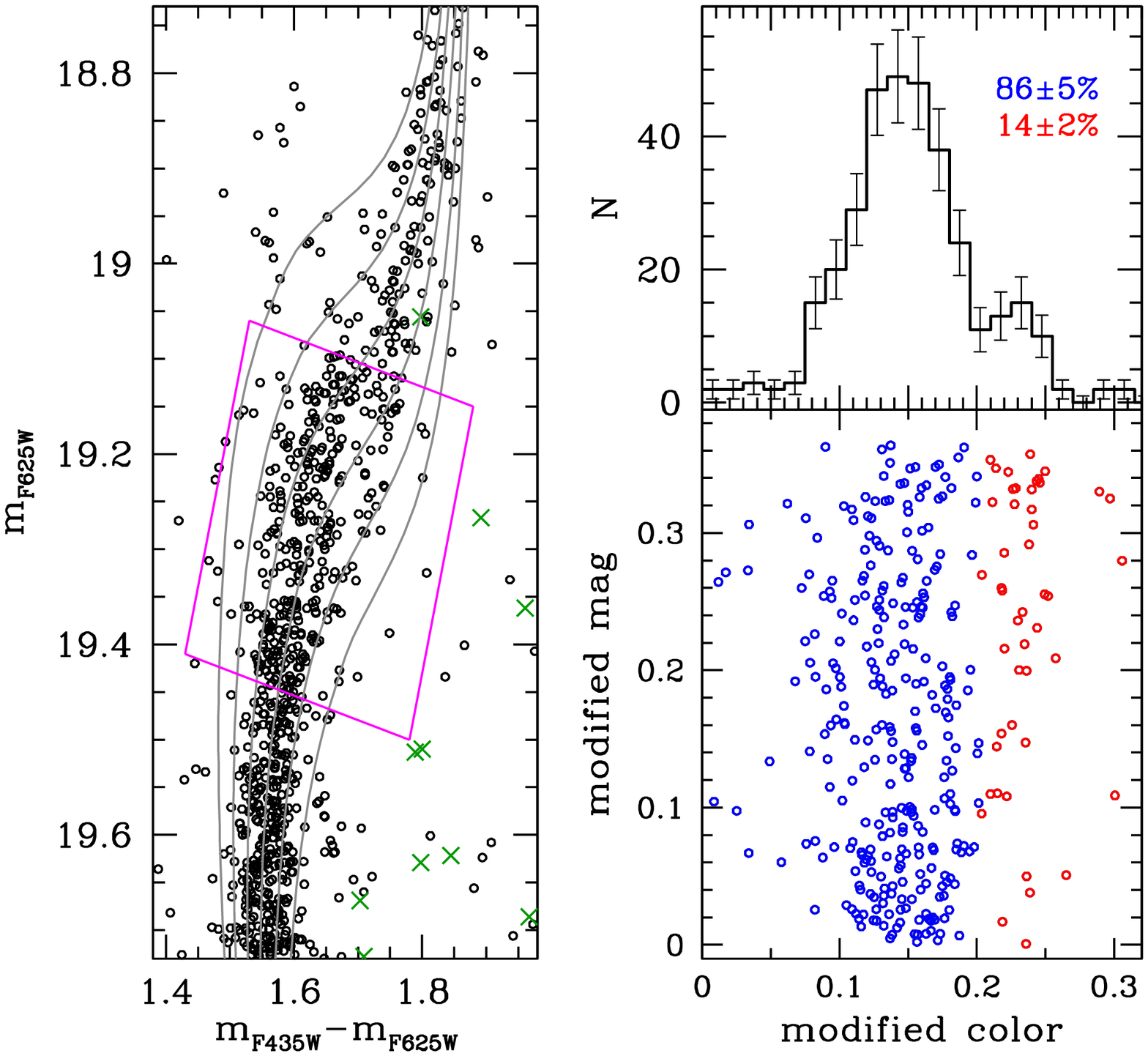}
\includegraphics[angle=0,scale=0.42]{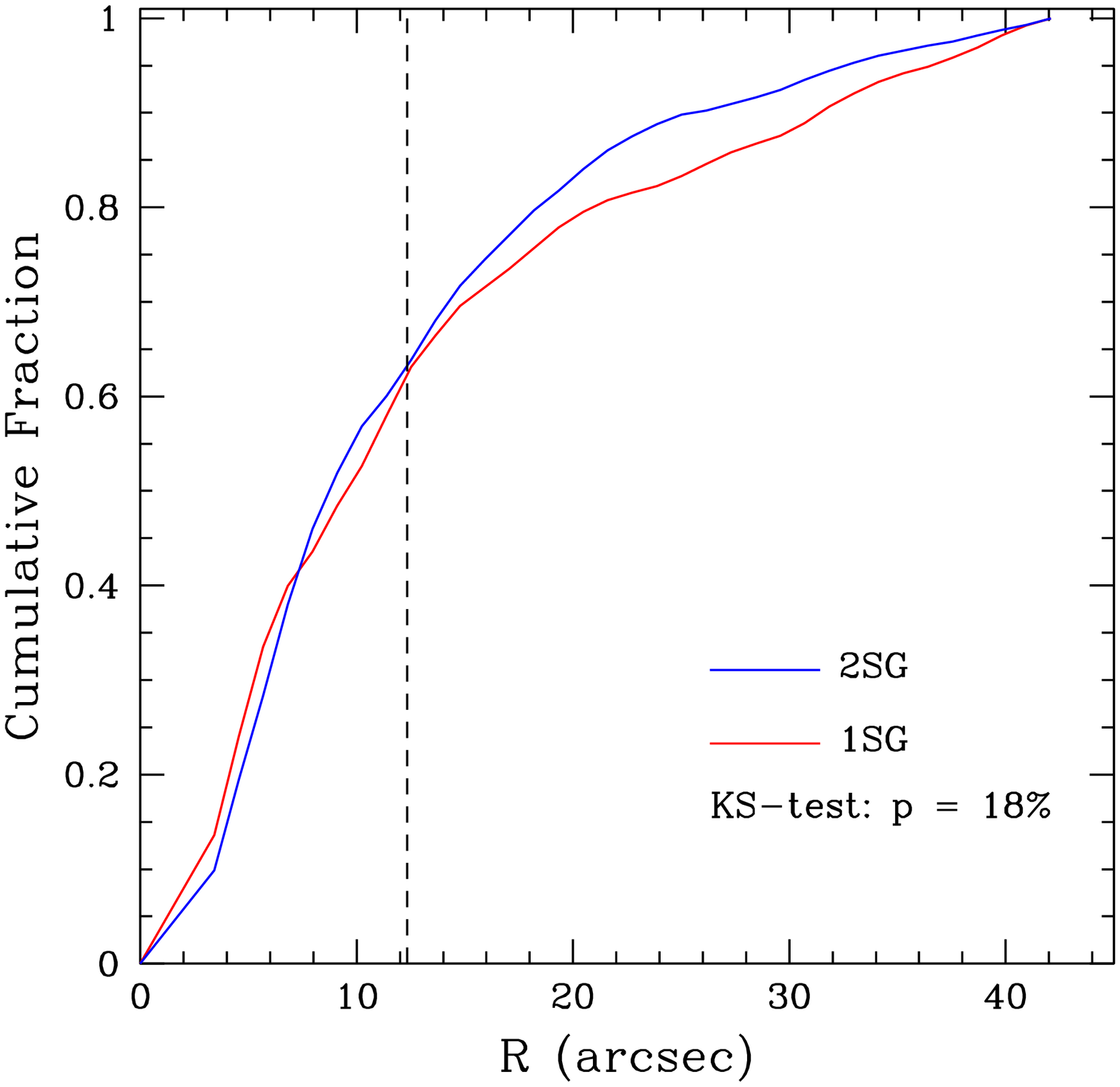}
\caption{Left panels: SGB of NGC\,6522 zoomed in, and ratio of candidates of first (red) to second (blue) stellar generations. 
DSED $\alpha$-enhanced isochrones for [Fe/H]=-1.01 and $Y=0.25$ with
10.0, 11.0, 12.0, 13.0, 14.0, and 15.0 Gyr are overplotted.
The expected residual field stars are also presented (green crosses).
Right panel: radial cumulative fraction distribution. The core radius (dashed line) and the K-S probability  are also given.}
\label{ratio} 
\end{figure*}

In order to detect a possible signature of MPs, we analyzed the SGB morphology. 
In the left panel of Figure \ref{ratio} it is shown that,
 in the $m_{\rm F625W}$ versus
$m_{\rm F435W}-m_{\rm F625W}$ proper-motion-cleaned CMD, the SGB of
NGC\,6522 is formed by two components: a more populated bright and blue SGB
(potentially associated with a second stellar generation -- 2G) 
 and a less populated faint and red SGB (a first stellar generation -- 1G).

An analysis of the ratio between the number of stars in each SGB
population and the total number of SGB stars is performed. The left
panels of Figure \ref{ratio} show the procedure adopted. First, the SGB sequence
is rectified using the procedure described in detail in \cite{Milone+09}
and \cite{Bellini+13}. The result is shown in the
bottom right panel of Figure \ref{ratio}.  The top right panel shows the abscissa
distribution (the bin size of the histogram is equal to 0.15).
Stars with 0.05 mag from the median of this distribution are the candidates
to a 1G population (red points).
By counting stars before and after this threshold, the fraction of
1G and 2G stars was determined.
The results indicate that $14\% \pm 2\%$ and $86\% \pm 5\%$ of the stars
belong to the 1G and 2G, respectively. Poisson uncertainties were assumed. 

The right panel of Figure \ref{ratio} presents the radial cumulative fraction distribution
of the 1G and 2G stars. The two distributions seem to agree inside the core radius;
however, at larger radii the 2G stars are more concentrated than the
1G stars.
A Kolmogorov--Smirnov (K-S) test provides a probability of 18\%  that the two samples were drawn
from the same underlying distribution.

To check whether the 1G sample is related to residual contaminating field stars we, compute
the expected number of such stars in the proper-motion-cleaned CMD taking into account
the spatial and kinematical information.
First, we defined a distance from the cluster center where the field stars are dominant
($R > R_{\rm{field}} = 100$ arcsec).
After that, we selected those with proper motions larger than 0.50 pixels (see Figure \ref{VPD_pm}).
Taking into account the areas in the sky and in the VPD that are covered by the cluster and field stars,
we found 110 stars that might be contaminating the cluster sample.
As can be seen in Figure \ref{ratio}, these stars are outside the box used to analyze the presence of
multiple populations, so they cannot explain the wide SGB observed in this cluster.
  
\begin{figure*}
\centering
\includegraphics[angle=0,scale=0.42]{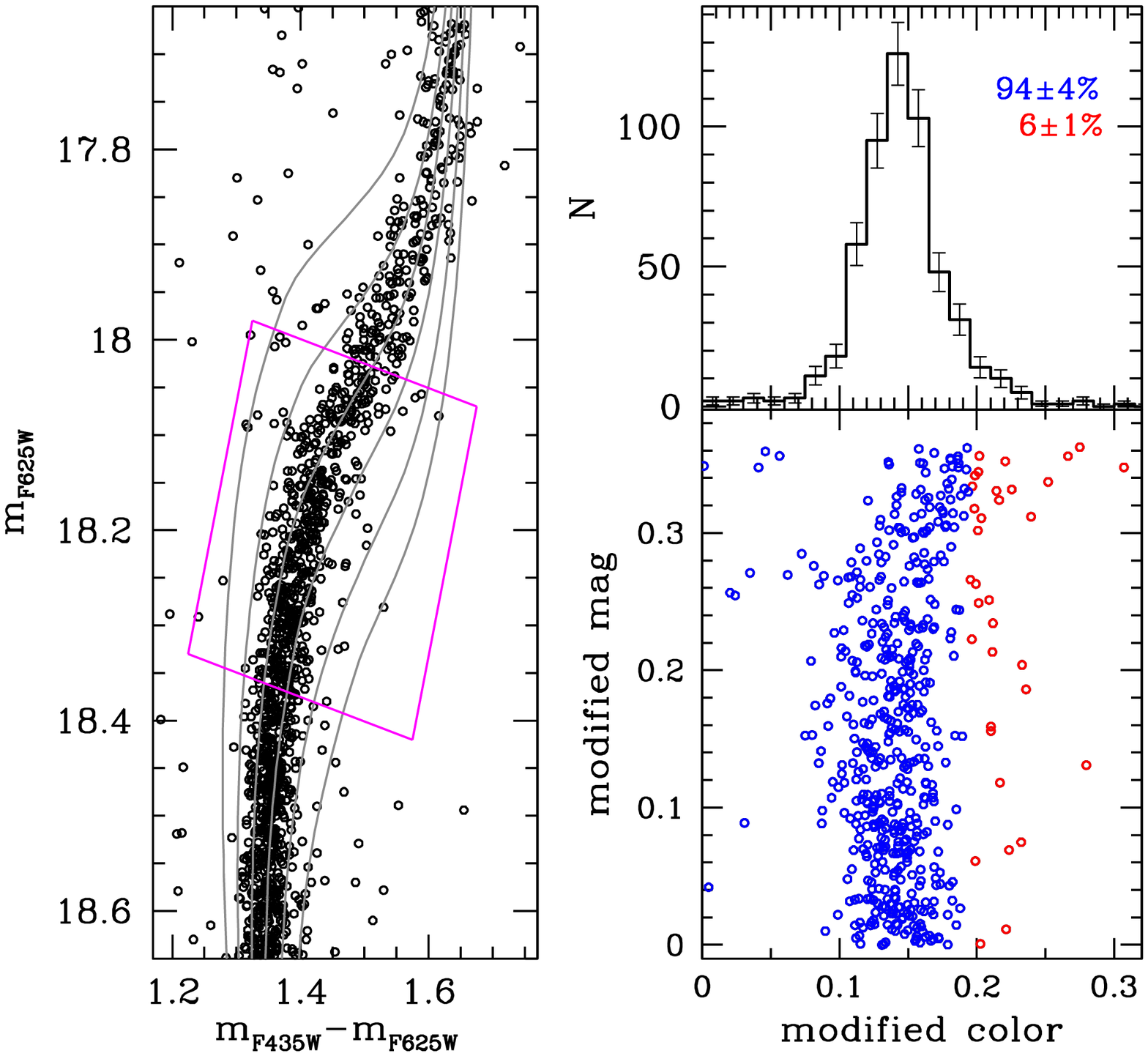}
\includegraphics[angle=0,scale=0.42]{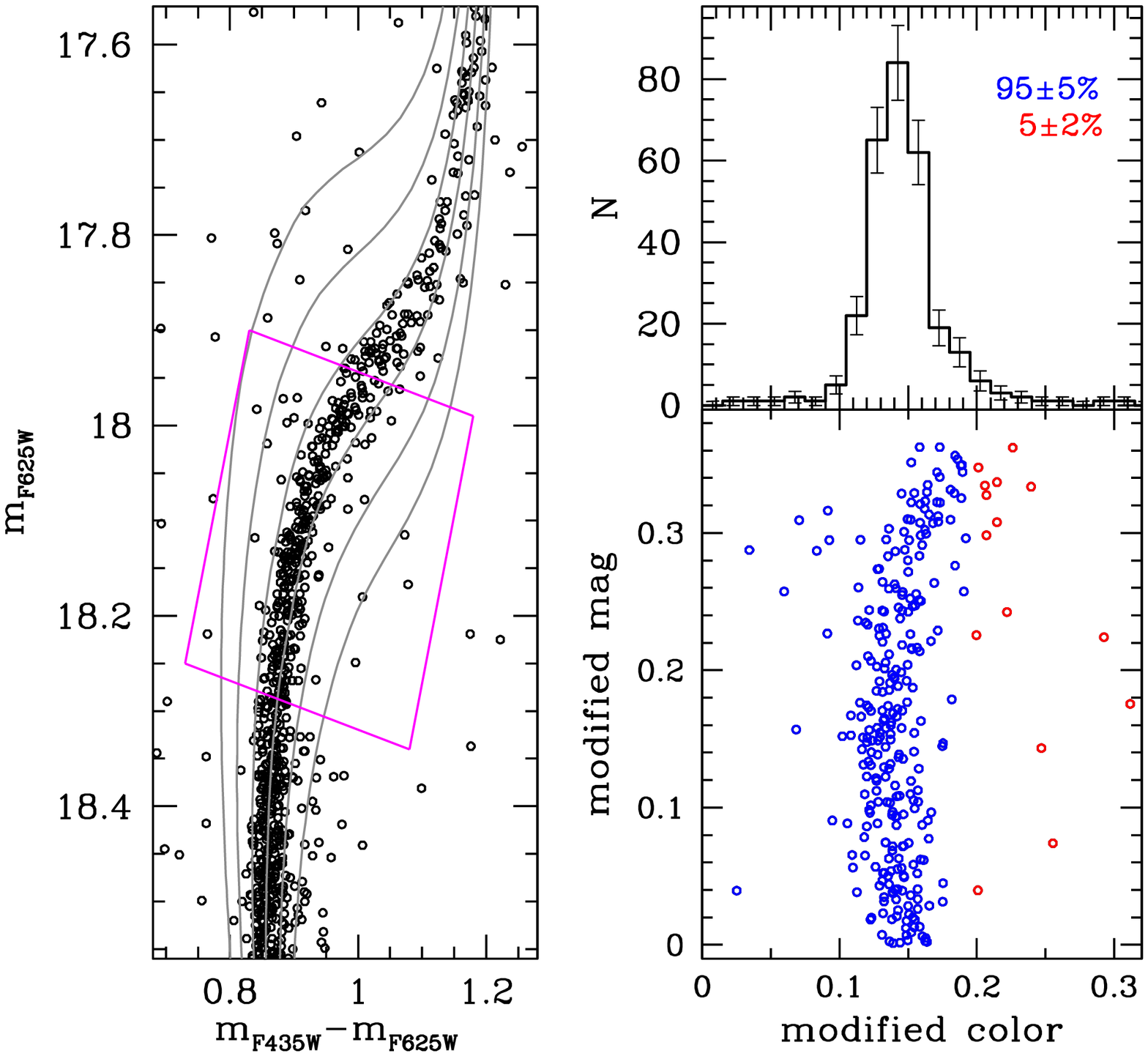}
\caption{Same as the left panels of Figure \ref{ratio}, but for NGC\,6626 (left panels) and NGC\,6362 (right panels).}
\label{ratio_other_clusters} 
\end{figure*}

In order to do a comparative study, we proceed with the same analysis
for the two other clusters.
These results are presented in Figure \ref{ratio_other_clusters}.
As can be seen in both panels, no more than $6 \% \pm 2\%$ of the stars in NGC\,6626 and NGC\,6362
could be associated with a 1G by the same method.
This result suggests that the 1G fraction of $14\% \pm 2\%$ for NGC\,6522 is a meaningful result,
as well as that the 1G proportion is probably a minimum value, and
that a higher proportion might be revealed when UV colors will
be available.
This is  expected because for NGC\,6362 we find a low 1G fraction, whereas
\cite{Milone+17} and \cite{Wagner-Kaiser+16}, analysing UV--optical \textit{HST} data,
determined higher values of 
$57.4\% \pm 3.5\%$ and $38.5^{+2.6}_{-2.5}\%$, respectively.

\subsection{Final remarks}

The combined constraints from MSTO-SGB-RGB and HB
 allowed us to conclude on final values of age, helium abundance,
distance, and reddening for the sample clusters. 
While the statistical comparisons of observed fiducial lines and theoretical
ones (Figure \ref{NGC6522_cmd_obs_vs_models_BaSTI}) appear to lead to the possibility
 of a higher-than-primordial helium abundances,
the HB and RR Lyrae levels, compared with theoretical models, 
imposed a normal helium content.

\begin{figure}
\centering
\includegraphics[angle=0,scale=0.42]{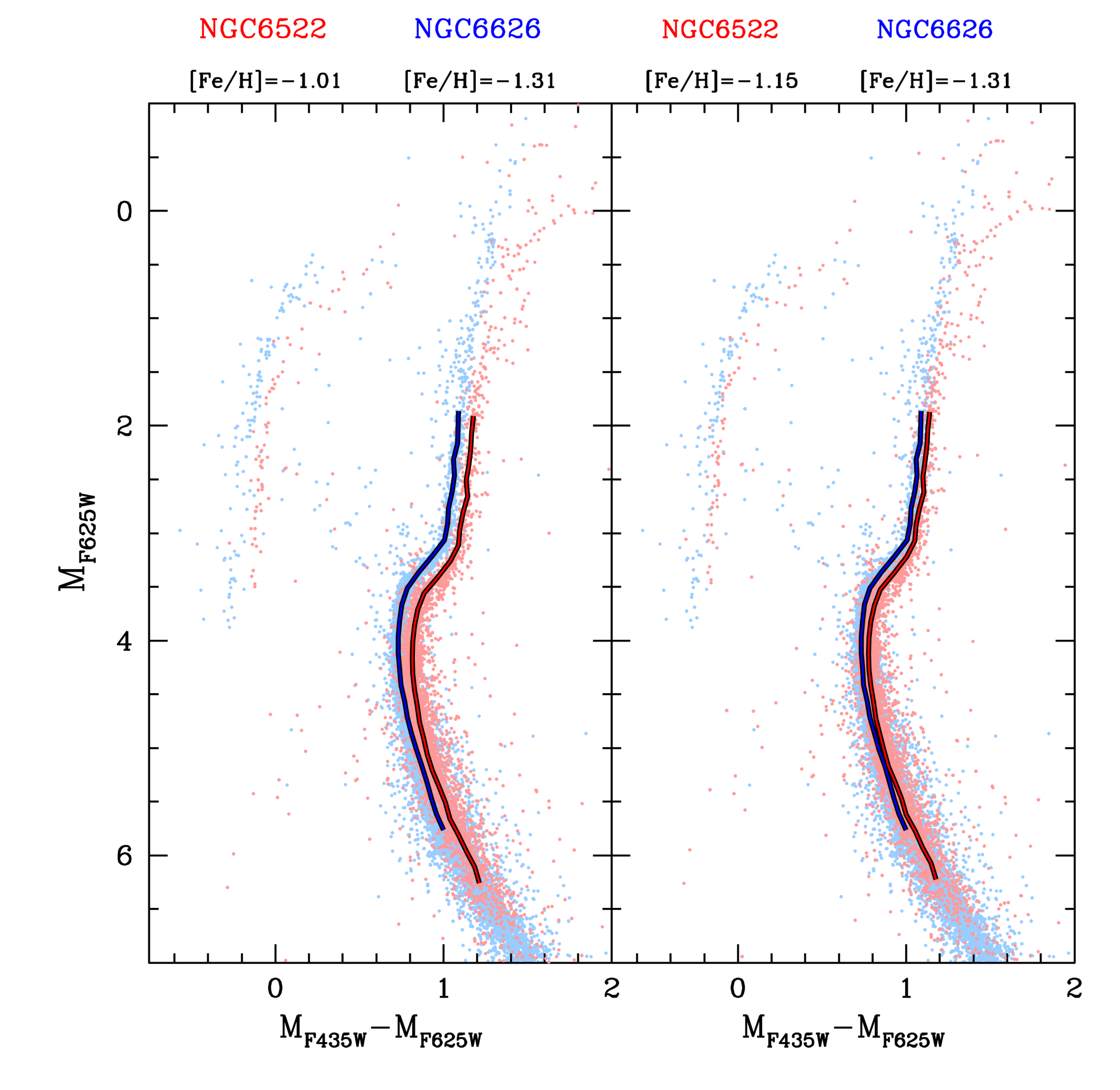}
\caption{CMDs and fiducial lines of NGC\,6522 and NGC\,6626
  in absolute magnitudes $M_{\rm{F625W}}$ vs. $M_{\rm{F435W}}-M_{\rm{F625W}}$.
  The conversion from observed magnitudes to the absolute ones takes into account
  the BaSTI stellar evolutionary models for the metallicities informed in each panel,
  as well as the distance moduli and reddening values that are consistent with the isochrone fits.
  A canonical helium abundance of $Y\sim 0.25$ was assumed.
}
\label{overplot1} 
\end{figure}

The observed CMDs of NGC\,6522 and NGC\,6626
are very similar (despite their differences in distance and reddening);
therefore, in the absolute magnitudes and colors they should almost coincide.
To verify this, in Figure \ref{overplot1} we show the CMDs and fiducial lines
of NGC\,6522 and NGC\,6626 transformed to absolute magnitudes and colors.  
Since the retrieved distance modulus and reddening do depend on the adopted metallicity,
in the left and right panels of Figure \ref{overplot1} we present the comparison between the two clusters
by adopting the parameters derived when assuming [Fe/H]$=-$1.0 and [Fe/H]$=-$1.15 for NGC\,6522,
respectively.
The reddening and distance modulus corrections change with metallicity and
lead the HB to fit better the sequences of NGC\,6626 in the latter.
With regard to the metallicity and abundances from high-resolution spectroscopy of NGC\,6522,
we should note two main issues:
(a) [$\alpha$/Fe]$<$+0.4 (Table \ref{tab_spec}), and in particular,
recalling that the bulk of stars are of a 2G generation,
oxygen should be lower,  whereas the models assume [$\alpha$/Fe]=+0.4;
(b) Na is enhanced, since the bulk of stars belong to a second generation. These abundance
variations can explain the need for a change in metallicity, with respect to NGC\,6626.
This shows the need, in the future, to have models taking into account
a range of element abundances, in particular of O, Na, Al, Mg, and
overall alpha-elements (Si, Ca, Ti, besides O, Mg).

\section{Conclusions}

We analyzed \textit{HST} proper-motion-cleaned CMDs of the bulge GCs
NGC\,6522 and NGC\,6626, together with the inner halo cluster NGC\,6362,
for comparison purposes. The three clusters have similar metallicities
($-1.3 \lesssim$ [Fe/H]$ \lesssim -1.0$) and alpha-enhancements ([$\alpha$/Fe]$\sim+0.4$).
They were observed in the same filters and (in the case of NGC\,6522 and NGC\,6626) in two epochs,
making it possible to select the likely cluster members.
Thus, these homogeneous and high-photometric-precision data 
allowed a detailed comparative analysis.

In this study we addressed the fundamental problem of deriving age,
reddening, and distance consistently, based on
statistical isochrone fitting and RR Lyrae analysis, including variable
helium content. 
The isochrone sets BaSTI and DSED were used, chosen because they offer
the needed parameters, in particular with regard to helium and alpha-element
abundances.


The use of BaSTI models leads to coeval ages of $\sim$ 13.0 Gyr 
for the three clusters,
or $\sim$ 14.0 Gyr if the atomic diffusion is neglected, independently of the
helium content. 
The results from DSED isochrones with canonical helium abundances
point to an age of $\sim$ 12.5 $\pm$ 0.5 Gyr for the clusters,
in very good agreement with ages given by
\cite{Dotter+10}, \cite{VandenBerg+13},
and \cite{Wagner-Kaiser+16} for NGC\,6362.
The fact that NGC\,6522 and NGC\,6626 have almost identical HB morphology, as shown
in Figure \ref{overplot1}, while at the same their metallicities differ
by $\sim$0.2--0.3 dex from high-resolution spectroscopic studies,
indicates that NGC\,6522 should be older than NGC\,6626.

The shape of the observed fiducial lines indicates some evidence of helium
enhancement for NGC\,6522 and NGC\,6626, but the average $V$ magnitudes of
RR Lyrae stars from Clement's catalog and the OGLE catalog tend to rule out this hypothesis,
since there is a good agreement between
these values and those expected from the recent empirical calibration
for the $\langle M_{V} \rangle$ Galactic RR Lyrae stars performed by the \cite{Gaia17}
(assuming the apparent distance moduli from our isochrone fits).
If the observed RR Lyrae stars in these clusters were helium enhanced by
$\Delta Y \sim 0.05$, they should be about 0.20 mag brighter on average, as predicted by BaSTI models.
Further inspection of HB morphology and new stellar evolutionary models will be the subject of a future work.

For the first time based on CMDs, we revealed that NGC\,6522 has
at least two stellar populations,
in proportions of approximately 86\% (2G) and 14\% (1G).
According to the study by \cite{Milone+17},
in his classification of type II GCs, the clusters that show clearly two stellar populations
in their so-called chromosome maps show a split subgiant branch (SGB) also in optical colors. NGC\,6522 
does show a split in the SGB in optical colors, whereas UV colors are not available for this cluster
for a definitive diagnostic as type II.
The type I clusters show a split SGB only in UV colors. This is the case of
NGC\,6362, where a  double stellar population is clearly detected with the UV colors, as shown by
\cite{Piotto+15} and \cite{Milone+17}. For NGC\,6626,
with the available optical data, there is no evidence so far of multiple stellar populations. 

Under the assumption that the total mass of NGC\,6522 is
 around half of that of NGC\,6362 \citep{Gnedin+Ostriker97}, 
NGC\,6522 establishes a new lower GC mass limit where both
photometric and spectroscopic signatures of multiple populations are present.

As a summary of main results, we emphasize that some parameters not available in the literature 
for these clusters were derived in the present work, in particular the age of NGC\,6626,
more precise ages for NGC\,6522,  and the multiple stellar populations of NGC\,6522. 

We conclude that the sample clusters, characterized by a moderate
 metallicity of [Fe/H]$\sim -1.0$ and enhanced $\alpha$-to-iron ratios,
 are among the oldest objects in the
Galaxy. This confirms that objects of this metallicity
 could have been formed from material previously enriched
as proposed in well-accepted scenarios of
bulge formation with fast chemical enrichment
\citep{Matteucci+Brocato90,Cescutti+08, Friaca+Barbuy17}.
It would be of great interest to have additional high-resolution
spectroscopy of stars in these clusters and to have NGC\,6522 
observed in UV filters to further probe its multiple stellar populations.

\acknowledgments

L.K., B.B., and E.B. acknowledge partial financial support from FAPESP, CNPq, and CAPES. 
D.N. and S.O. acknowledge support by the Universit\`a degli Studi di Padova
Progetto di Ateneo CPDA141214, ``Towards Understanding Complex Star Formation
in Galactic GCs.'' 
S.C. acknowledges the financial support by PRIN-INAF2014 (PI: S. Cassisi)
and the Economy and Competitiveness Ministry of the Kingdom of Spain (grant AYA2013-42781-P).
M.L. recognizes partial support by PRIN-INAF 2014, ``The Kaleidoscope of stellar populations in Galactic GCs with Hubble Space Telescope.''R.G.V. acknowledges the support from FAPESP (grant 2012/20364-4).

%

\vspace{5mm}
\facilities{\textit{HST}(ACS,WFPC2).}

 \end{document}